\documentstyle[aps,floats,prl]{revtex}
\input epsf
\newcommand{\el}{l}
\newcommand{\bn}{\hat{\bf n}}
\newcommand{\even}{\rm{even}}
\newcommand{\odd}{\rm{odd}}
\newcommand{\bq}{\hat{\bf q}}
\newcommand{\tot}{{\rm tot}}

\newcommand{\trispect}{{\cal T}}
\newcommand{\bl}{{\bf l}}
\newcommand{\Ylm}[1]{Y_{\el_#1}^{m_#1}}
\newcommand{\Ylmp}[1]{Y_{\el_#1}^{m_#1'}}
\newcommand{\wj}[6]{\left(
                           \begin{array}{ccc}
        \! #1\! & #2\!  & #3\!  \\
        \! #4\! & #5\!  & #6\!
                           \end{array}
                   \right)}
\newcommand{\wsixj}[6]{\left\{
                           \begin{array}{ccc}
         #1 & #2  & #3  \\
         #4 & #5  & #6
                           \end{array}
                   \right\}}
\newcommand{\ApJL}{ApJ Lett.}
\newcommand{\ApJ}{ApJ}
\newcommand{\PRL}{Phys. Rev. Lett.}
\newcommand{\PRD}{Phys. Rev. D}

\newcommand{\aut}[2]{{#2.\ #1,}}
\newcommand{\laut}[2]{{#2.\ #1,}}
\newcommand{\refs}[6]{#2, #3  {#4} (#5).}

\newcommand{\mybib}[2]{\bibitem{#2}}

\begin{document}

\title{The Angular Trispectrum of the CMB}

\author{Wayne Hu}

\address{5640 S. Ellis Ave, University of Chicago, Chicago, IL 60637\\
Revised \today}
\maketitle
\begin{abstract}
We study the general properties of the CMB temperature four-point 
function, specifically its harmonic analogue the angular trispectrum,  
and illustrate its utility in finding optimal quadratic statistics through
the weak gravitational lensing effect. 
We determine the general form of the trispectrum, under the assumptions of
rotational, permutation, and parity invariance, its estimators on the sky,
and their Gaussian noise properties.  The signal-to-noise in the
trispectrum can be highly configuration dependent and any 
quadratic statistic used to compress the information to a manageable two-point
level must be carefully chosen.  Through a systematic study,
we determine that for the case of lensing a specific statistic,
the divergence of a filtered temperature-weighted temperature-gradient map,
contains the maximal signal-to-noise and reduces the variance of
estimates of the large-scale convergence power spectrum by over an
order of magnitude over previous gradient-gradient techniques.  
The total signal-to-noise for lensing with the Planck satellite is 
of order 60 for a $\Lambda$CDM cosmology.
\end{abstract}

\section{Introduction}

The power spectra or two-point correlations of 
Cosmic Microwave Background (CMB) temperature and polarization
anisotropies are widely recognized as a gold mine of 
information on cosmology.  These spectra in fact contain all
of the information embedded in the CMB if the underlying fluctuations
are Gaussian distributed.  
Nonetheless even if the initial density fluctuations are Gaussian, 
non-Gaussianity
in the CMB temperature fluctuations will be generated by non-linear
processes.  These generally are associated with the secondary anisotropies
that are imprinted as the photons propagate through the large-scale
structure of the Universe from the epoch of recombination.  

Secondary signatures in the three-point correlation of temperature 
anisotropies have recently received much attention 
\cite{GolSpe99,SelZal98i,CooHu00a}
following early pioneering work on intrinsic correlations in
the initial conditions \cite{early,Luo94}.
The four-point correlation and its harmonic analogue the trispectrum
has received considerably less attention despite the fact that
it directly controls the noise properties of estimators of the power spectrum.
In particular, an all-sky treatment of the trispectrum that incorporates
the full rotational symmetry properties of the trispectrum has been lacking
in the literature (c.f. \cite{Zal00}).  Exploitation of the
symmetry properties can assist in the isolation of the physical mechanisms
underlying the generation of the trispectrum as we shall see.

In this paper, we establish the framework needed to study the trispectrum 
on the full sky. 
We begin in \S \ref{sec:symmetries} with a discussion of the symmetry
properties of the $n$-point function on the sky, with an
emphasis on the 4-point function, and their implications for the
general form of the harmonic spectra.  We consider estimators
of the trispectrum and their noise properties in \S \ref{sec:estimatorssn},
and the trispectrum-based power spectra of quadratic statistics in
\S \ref{sec:collapsedtri}.  Calculational techniques and relationships 
to the flat-sky formalism are given in two Appendices. 
In \S \ref{sec:lensing}, we consider
the specific case of the trispectrum generated by weak gravitational
lensing of CMB photons by the large-scale structure of the Universe
and show that there exists a quadratic statistic that optimally
recovers the projected gravitational potential power spectrum  
(or convergence) on large scales.  We conclude \S \ref{sec:discussion}.

\section{Symmetries}
\label{sec:symmetries}

In this section, we derive the requirements that rotational, permutation, 
and parity symmetry impose on the $n=(2, 3, 4)$-point correlation functions
on the sphere and their spherical harmonic analogues: the power spectrum,
bispectrum and trispectrum.  We begin with general considerations for
the $n$-point function in \S \ref{sec:general}, review the implications for
the power spectrum and bispectrum in \S \ref{sec:powerbi}, and derive
the consequences for the trispectrum in \S \ref{sec:trispectrum}.  
In \S \ref{sec:enforce}, we show how to construct trispectra with the
required symmetry properties.

\subsection{General Considerations}
\label{sec:general}

We begin by requiring statistical isotropy of the $n$-point correlation
function on the sphere and its harmonic analogue
\begin{equation}
\langle \Theta(\bn_1) \ldots \Theta(\bn_n) \rangle =\sum_{\el_1 \ldots \el_n}
\sum_{m_1 \ldots m_n} \langle \Theta_{\el_1 m_1} \ldots \Theta_{\el_n m_n} \rangle
\Ylm{1}(\bn_1)\ldots \Ylm{n}(\bn_n) \,.
\end{equation}
Statistical isotropy demands that the $n$-point function is
invariant under an arbitrary rotation $R$ whose action on
a spherical harmonic is expressed in terms of the Wigner-D
function
\begin{equation}
R[\Ylm{{}}(\bn)] = \sum_{m'} D_{m' m}^\el (\alpha,\beta,\gamma) 
\Ylmp{{}} (\bn)\,,
\end{equation}
where $\alpha$, $\beta$ and $\gamma$ are the Euler angles of the
rotation.  To obey rotational invariance the harmonics must obey the 
relation
\begin{equation}
\langle \Theta_{\el_1 m_1} \ldots \Theta_{\el_n m_n} \rangle
= \sum_{m_1' \ldots m_n'} 
\langle \Theta_{\el_1 m_1'} \ldots \Theta_{\el_n m_n'} \rangle
	D_{m_1 m_1'}^{\el_1}  \ldots
	D_{m_n m_n'}^{\el_n}\,,
\end{equation}
for all $\alpha$, $\beta$, and $\gamma$.
The reduction of this relation proceeds as follows.  Each pair
of rotation matrices may be coupled into a single rotation
via the group multiplication property (or equivalently the
addition of angular momentum)
\begin{eqnarray}
	D_{m_1 m_1'}^{\el_1}  
	D_{m_2 m_2'}^{\el_2} 
	&=& \sum_{L M M'} \wj{\el_1}{\el_2}{L}
			  {m_1}{m_2}{-M}
	               \wj{\el_1}{\el_2}{L}
			  {m_1'}{m_2'}{-M'} 
	(2L+1)(-1)^{M+M'} D_{M M'}^{L} \,.
\label{eqn:multiplication}
\end{eqnarray}
When the product is reduced a pair of $D$ matrices, one seeks the
form of the harmonic $n$-point function that reduces the pair to
the orthogonality condition for rotations
\begin{eqnarray}
	\sum_m (-1)^{m_2-m  } D_{m_1 m}^{\el_1} D_{-m_2 -m}^{\el_1} 
		&=& \delta_{m_1 m_2} \,,
\label{eqn:unitarity}
\end{eqnarray}
which is valid for an arbitrary rotation.  The indices can then be
permuted to find alternate orderings of the pairings.

Invariance under a parity transformation which takes
$\bn \rightarrow -\bn$
\begin{equation}
Y_\el^m \rightarrow (-1)^{\el} Y_\el^m\,,
\end{equation}
would require that 
\begin{equation}
\sum_{i=1}^n \el_i = {\rm even.}
\end{equation}
Reality of the underlying $\Theta$ field and the fact that
\begin{equation}
Y_\el^{m*} = (-1)^m Y_\el^{-m}\,,
\end{equation}
requires that
\begin{equation}
\Theta_\el^{m*} = (-1)^m \Theta_\el^{-m}\,.
\end{equation}

\subsection{Power Spectrum and Bispectrum}
\label{sec:powerbi}

For the 2-point function there is only one step.  The reduction of equation
(\ref{eqn:unitarity}) requires the form
\begin{equation}
\langle \Theta_{\el_1 m_1} \Theta_{\el_2 m_2} \rangle = 
	\delta_{\el_1 \el_2} \delta_{m_1 -m_2} (-1)^{m_1} C_{\el_1}\,.
\end{equation}
For the 3-point function one first collapses one product of rotation matrices
leaving
\begin{eqnarray}
\langle \Theta_{\el_1 m_1} \ldots \Theta_{\el_3 m_3} \rangle &=&
\sum_{m_1' \ldots m_3'} 
\langle \Theta_{\el_1 m_1'} \ldots \Theta_{\el_3 m_3'} \rangle 
 \sum_{L M M'} (2L+1)(-1)^{M+M'}\\
&&\times   \wj{\el_1}{\el_2}{L}
			  {m_1}{m_2}{-M}
	               \wj{\el_1}{\el_2}{L}
			  {m_1'}{m_2'}{-M'} 
			D_{M M'}^{L} D_{m_3 m_3'}^{\el_3} \,.
			\nonumber
\end{eqnarray}	
In order to reduce this relation 
to the orthogonality condition Eqn.~(\ref{eqn:unitarity}), 
the sum over $m'$ of
the three-point function must be proportional to 
$\delta_{L \el_3} \delta_{M' -m_3'}$.  Recalling the identity
\begin{equation}
\sum_{m_1 m_2}  
		\wj{\el_1}{\el_2}{L}
	 	   {m_1}{m_2}{M}
                \wj{\el_1}{\el_2}{L'}
	 	   {m_1}{m_2}{M'} = {\delta_{L L'} \delta_{M M'} \over 2L+1}\,,
\end{equation}
we can obtain the desired relation if the $m$ dependence of the
3-point function obeys
\begin{equation}
\langle \Theta_{\el_1 m_1}\ldots \Theta_{\el_3 m_3} \rangle =
		\wj{\el_1}{\el_2}{\el_3} 
                   {m_1}{m_2}{m_3} B_{\el_1 \el_2 \el_3}\,.
\end{equation}

\subsection{Trispectrum}
\label{sec:trispectrum}

The form of the 4-point function follows the same steps except that
we use the group multiplication properties to pair say ($\el_1$,$\el_2$)
and ($\el_3$,$\el_4$) leading to the condition
\begin{eqnarray}
\langle \Theta_{\el_1 m_1}\ldots \Theta_{\el_4 m_4}\rangle &=&
\sum_{m_1'\ldots m_4'} 
\sum_{L_{12} M_{12} M'_{12}}  
\!\!\!\!\!\!
(2L_{12}+1)(-1)^{M_{12}+M_{12}'}
		\wj{\el_1}{\el_2}{L_{12}}
			  {m_1}{m_2}{-M_{12}}
	               \wj{\el_1}{\el_2}{L_{12}}
			  {m_1'}{m_2'}{-M_{12}'}
			D_{M_{12} M'_{12} }^{L_{12}} \\
&&\times \!\!\!\!\!\! \sum_{L_{34} M_{34} M'_{34}}   
\!\!\!\!\!\!
(2L_{34}+1)(-1)^{M_{34}+M_{34}'}
\wj{\el_3}{\el_4}{L_{34}}
			  {m_3}{m_4}{-M_{34}}
	               \wj{\el_3}{\el_4}{L_{34}}
			  {m_3'}{m_4'}{-M_{34}'} 
			D_{M_{34} M'_{34} }^{L_{34}} 
\langle \Theta_{\el_1 m_1'}\ldots \Theta_{\el_4 m_4'} \rangle \nonumber
\,.
\end{eqnarray}
The same reasoning that led to the choice of the form of the three-point
function implies that the following form is a solution 
\begin{equation}
\langle \Theta_{\el_1 m_1}\ldots \Theta_{\el_4 m_4}\rangle=
	\sum_{LM} 
	               \wj{\el_1}{\el_2}{L}
			  {m_1}{m_2}{-M} 
	               \wj{\el_3}{\el_4}{L}
			  {m_3}{m_4}{M}  (-1)^M 
			Q^{\el_1 \el_2}_{\el_3 \el_4}(L)\,.
\label{eqn:triform}
\end{equation}
Geometrically, $Q^{\el_{1}\el_{2}}_{\el_{3}\el_{4}}(L)$
represents a quadrilateral 
composed of sides with length $\el_{1} \ldots \el_{4}$.
The index $L$ represents one of the diagonals of the quadrilateral
and is also the shared third side of the two triangles formed by the
corresponding pairs of sides.  The Wigner-$3j$ symbols in 
Eqn.~(\ref{eqn:triform}) ensure that the triangle inequalities
are satisfied.  For this reason we will often refer to
a set ($\el_{1}$, $\el_{2}$, $\el_3$, $\el_4$, $L$) as
a given ``configuration'' of the quadrilateral.

The two other unique pairings
of the indices,  $(\el_{1},\el_{3})$ and $(\el_{1},\el_{4})$, yield
alternate representations of the 4-point function.  These are 
not independent since all three couplings yield complete sets according
to the theory of the addition of angular momenta.   The alternate 
representations are constructed as linear combinations of the
$(\el_{1},\el_{2})$ representation with weights given by the Wigner-6$j$
recoupling coefficients (see Appendix)
\begin{eqnarray}
Q^{\el_1 \el_3}_{\el_2 \el_4}(L) &=&
	        \sum_{L'} (-1)^{\el_2 + \el_3} (2L+1)
		\wsixj{\el_1}{\el_2}{L'}{\el_4}{\el_3}{L} 
	        Q^{\el_1 \el_2}_{\el_3 \el_4}(L')\,,\nonumber\\
Q^{\el_1 \el_4}_{\el_3 \el_2}(L) &=&
	        \sum_{L'} (-1)^{L+L'} (2L+1)
		\wsixj{\el_1}{\el_2}{L'}{\el_3}{\el_4}{L} 
	        Q^{\el_1 \el_2}_{\el_3 \el_4}(L')\,,
\label{eqn:pairsymmetry}
\end{eqnarray}
where we have used Eqn.~(\ref{eqn:w3jw6j}) to project one coupling
scheme onto another.

Symmetry with respect to the $4!/3 =8$ remaining permutations (2 orderings
of the pairs, 4 orderings within the pairs) requires that
\begin{eqnarray}
Q^{\el_1 \el_2}_{\el_3 \el_4}(L) &=& 
		(-1)^{\Sigma_U} Q^{\el_2 \el_1}_{\el_3 \el_4}(L) =
		(-1)^{\Sigma_L} Q^{\el_1 \el_2}_{\el_4 \el_3}(L) =
		Q^{\el_3 \el_4}_{\el_1 \el_2}(L)\,, 
\end{eqnarray}
where $\Sigma_U = \el_1+\el_2+L$ and $\Sigma_L=\el_3+\el_4+L$.
If the four-point function is parity invariant then
\begin{equation}
Q^{\el_1 \el_2}_{\el_3 \el_4}(L) =
Q^{\el_2 \el_1}_{\el_4 \el_3}(L) \,.
\end{equation}
We shall show how to construct trispectra that obey these properties
in the next section.

Finally it is useful to separate the contributions from
the unconnected or Gaussian piece and the connected or trispectrum piece
\begin{equation}
Q^{\el_1 \el_2}_{\el_3 \el_4}(L) = 
G^{\el_1 \el_2}_{\el_3 \el_4}(L) +  
T^{\el_1 \el_2}_{\el_3 \el_4}(L) \,,
\end{equation}
where
\begin{eqnarray}
G^{\el_1 \el_2}_{\el_3 \el_4}(L)
&=& (-1)^{\el_1+\el_3} \sqrt{(2\el_1+1)(2\el_3+1)} C_{\el_1} C_{\el_3}
	\delta_{\el_1 \el_2} \delta_{\el_3 \el_4} \delta_{L0}
	+(2L+1) C_{\el_1} C_{\el_2} \left[
(-1)^{\el_2+\el_3+L} \delta_{\el_1 \el_3} \delta_{\el_2 \el_4} 
+ \delta_{\el_1 \el_4} \delta_{\el_2 \el_3}\right]\,.
\label{eqn:gaussian}
\end{eqnarray}

\subsection{Enforcing Symmetries}
\label{sec:enforce}

The symmetries of the trispectrum described above may be enforced by
the following construction.  First describe the four-point function
by a form that is explicitly symmetric in the three unique pairings
\begin{eqnarray}
\langle \Theta_{\el_1 m_1}\ldots \Theta_{\el_4 m_4}\rangle_c & = &
			\sum_{LM} 
			P^{\el_1 \el_2}_{\el_3 \el_4}(L)
	               \wj{\el_1}{\el_2}{L}
			  {m_1}{m_2}{-M} 
	               \wj{\el_3}{\el_4}{L}
			  {m_3}{m_4}{M}  (-1)^M 
+(\el_2  \leftrightarrow \el_3) + (\el_2 \leftrightarrow \el_4)\,,
			\label{eqn:permform} 
\end{eqnarray}
where $c$ denotes the fact that we have removed the Gaussian piece
of Eqn.~(\ref{eqn:gaussian}).
The two latter pairings can be projected
onto the  $(\el_{1},\el_{2})$ basis with the help of the
Wigner-6$j$ symbol to give
\begin{eqnarray}
T^{\el_1 \el_2}_{\el_3 \el_4}(L) &=&
			P^{\el_1 \el_2}_{\el_3 \el_4}(L)
+ (2L+1)\sum_{L'} \left[ (-1)^{\el_2+\el_3} 
\wsixj{\el_1}{\el_2}{L}{\el_4}{\el_3}{L'} 
P^{\el_1 \el_3}_{\el_2 \el_4}(L')
+(-1)^{L+L'} 
\wsixj{\el_1}{\el_2}{L}{\el_3}{\el_4}{L'} 
			P^{\el_1 \el_4}_{\el_3 \el_2}(L')\right]\,.
\end{eqnarray}
Within the three unique pairings, there are 4 permutations of the
ordering implying that $P$ is constructed as 
\begin{eqnarray}
P^{\el_{1} \el_{2}}_{\el_{3} \el_{4}}(L) &=&
\trispect^{\el_{1} \el_{2}}_{\el_{3}\el_{4}}
+ (-1)^{\Sigma_{U}}
\trispect^{\el_{2} \el_{1}}_{\el_{3}\el_{4}}
+ (-1)^{\Sigma_{L}}
\trispect^{\el_{1} \el_{2}}_{\el_{4}\el_{3}}
+ (-1)^{\Sigma_{U} + \Sigma_{L}}
\trispect^{\el_{2} \el_{1}}_{\el_{4}\el_{3}}\,.
\label{eqn:reducedtrispect}
\end{eqnarray}
The reduced function $\cal T$ underlying the trispectrum is an arbitrary
function of its arguments except that it must be symmetric against
exchange of its upper and lower indices
\begin{equation}
\trispect^{\el_{1} \el_{2}}_{\el_{3} \el_{4}}(L) =
 \trispect^{\el_{3} \el_{4}}_{\el_{1} \el_{2}}(L)\,,
\end{equation}
and if parity invariant obeys 
\begin{equation}
\trispect^{\el_{1} \el_{2}}_{\el_{3} \el_{4}}(L) =
 \trispect^{\el_{2} \el_{1}}_{\el_{4} \el_{3}}(L)\,.
\end{equation}
This then completes the enforcing of the rotation, permutation and parity
symmetries of the trispectrum.

\section{Estimators and Signal-to-Noise}
\label{sec:estimatorssn}

We show in \S \ref{sec:estimators} that the fundamental estimator
of the trispectrum involves a weighted sum over the multipole moments
in a given quadruplet of harmonics.  
These estimators have well defined noise properties
as derived in \S \ref{sec:signaltonoise} which can be used to 
calculate the theoretical signal-to-noise in the trispectrum. 
A non-vanishing trispectrum can on the other hand decrease the signal-to-noise
in the power spectrum by introducing a covariance between its
estimators as shown in \S \ref{sec:pscovariance}.

\subsection{Estimators}
\label{sec:estimators}

From the orthogonality properties of the Wigner-3$j$ symbol, one can invert
the relationship for the four-point spectrum to form the estimator
\begin{eqnarray}
\hat T^{\el_1 \el_2}_{\el_3 \el_4}(L) &=& (2L+1) \sum_{m_1 m_2 m_3 m_4 M}
(-1)^M  
		\wj{\el_1}{\el_2}{L} {m_1}{m_2}{M} 
		\wj{\el_3}{\el_4}{L} {m_3}{m_4}{-M}
		\langle 
		\Theta_{\el_1 m_1} 
\ldots \Theta_{\el_4 m_4}
	        \rangle\, - \hat G^{\el_1 \el_2}_{\el_3 \el_4},
\label{eqn:estimator}
\end{eqnarray}
where the estimator for the Gaussian piece is constructed out of
those for the power spectrum.   Note that for configurations whose
$\el$ sides are not equal in pairs, 
the Gaussian piece vanishes and the sum over
$m$'s of the spherical harmonic coefficients is an unbiased estimator
of the trispectrum.

We can alternately
form an  estimator of particular configurations of the 
trispectrum directly from the sky map itself without an
explicit expansion in spherical harmonics.  Following Spergel \& Goldberg 
\cite{SpeGol99}, let us define a new set of sky maps weighted in rings
centered around a point $\bq$:
\begin{equation}
e_\el(\bq) = \sqrt{2\el+1 \over 4\pi} \int d\bn \Theta(\bn) P_\el(\bn \cdot \bq)\,.
\end{equation}
Expanding the Wigner-3$j$ symbols in terms of spherical harmonics and using
the addition theorem, we obtain
\begin{eqnarray}
&& \wj{\el_1}{\el_2}{L}{0}{0}{0}
\wj{\el_3}{\el_4}{L}{0}{0}{0} \left[
\hat T^{\el_1 \el_2}_{\el_3 \el_4}(L) 
+ \hat G^{\el_1 \el_2}_{\el_3 \el_4}(L) \right]
= 
\nonumber\\&&\quad 
(2L+1) 
\int {d \bq_a \over 4\pi} 
\int {d \bq_b \over 4\pi} 
e_{\el_1}(\bq_a) e_{\el_2}(\bq_a)
e_{\el_3}(\bq_b) e_{\el_4}(\bq_b) P_L(\bq_a \cdot \bq_b)\,.
\end{eqnarray}
Since the Wigner-3$j$ symbol vanishes if $\el_1+\el_2+L =$ odd,
this expression can only be used to estimate even terms.

To measure all configurations of the trispectrum is, needless to
say, a daunting task.  Aside from the computational expense, one
must also treat complications associated with estimators of
harmonics on a fraction of the sky.  Even for an all-sky CMB experiment, 
the removal of galactic foregrounds will limit the data to a smaller
fraction of the sky $f_{\rm sky}$. 

\subsection{Signal-to-Noise}
\label{sec:signaltonoise}

Returning to the estimator of Eqn.~(\ref{eqn:estimator}), 
one can calculate the Gaussian {\it noise} variance of the estimator,
\begin{eqnarray}
\langle 
\hat T^{\el_1 \el_2*}_{\el_3 \el_4}(L) 
\hat T^{\el_1 \el_2}_{\el_3 \el_4}(L')  \rangle
= (2 L+1) 
	\delta_{L L'} 
	C_{\el_1}^\tot
	C_{\el_2}^\tot
	C_{\el_3}^\tot
	C_{\el_4}^\tot\,,
\end{eqnarray}
if no two $\el$'s are equal.
Here $C_\el^\tot$ is the sum of all contributions to the power spectrum
including the intrinsic CMB fluctuations, 
instrumental noise and residual foreground 
contamination.

From the permutation properties of $Q$ (or $T$) in  Eqn.~(\ref{eqn:pairsymmetry}), 
the full covariance of the
estimators then becomes
\begin{eqnarray}
{\langle \hat T^{\el_1 \el_2*}_{\el_3 \el_4}(L) 
\hat T^{\el_1' \el_2'}_{\el_3' \el_4'}  (L')  \rangle \over
	(2L+1) C_{\el_1}^\tot
	C_{\el_2}^\tot
	C_{\el_3}^\tot
	C_{\el_4}^\tot}
&=& \delta_{L L'} 
	\delta^{12}_{34}
 + (2L'+1) \left[ (-1)^{\el_2+\el_3} 
      \wsixj{\el_1}{\el_2}{L}{\el_4}{\el_3}{L'} 
	\delta^{13}_{24}
+ (-1)^{L+L'}
       \wsixj{\el_1}{\el_2}{L}{\el_3}{\el_4}{L'} 
	\delta^{14}_{32} \right]\,,
\label{eqn:covariance}
\end{eqnarray}
if no two $\el$'s in the primed and unprimed sets are equal.
Here
\begin{equation}
\delta^{ab}_{cd} = [
	\delta_{\el_1 \el_a} \delta_{\el_2 \el_b} +
	(-1)^{\Sigma_U} 
	\delta_{\el_1 \el_b} \delta_{\el_2 \el_a}]
	[\delta_{\el_3 \el_c} \delta_{\el_4 \el_d} +
	(-1)^{\Sigma_L}
	\delta_{\el_3 \el_d} \delta_{\el_4 \el_c}]
	+ [{a \leftrightarrow c}][{b \leftrightarrow d}]\,,
\end{equation}
accounts for the permutations within the three fundamental pairings.	
Recall that $\Sigma_U=\el_1+\el_2+L$ and $\Sigma_L=\el_3+\el_4+L$
The two terms involving the Wigner-6$j$ symbol reflect the
fact that alternate pairings of the indices supply redundant information
in both the signal and the noise.  

If any two $\el$'s are equal, then the covariance has extra terms 
associated with the internal pairings in the primed and unprimed
sets.  Based on the fundamental relation
\begin{equation}
{\langle \hat T^{\el_1 \el_1}_{\el_3 \el_4}(L) 
\hat T^{\el_1' \el_1'}_{\el_3' \el_4'}  (L')  \rangle}
= (-1)^{\el_1 + \el_1'}  
	\delta_{L0} \delta_{L'0} 
	\sqrt{( 2 \el_1+1)( 2 \el_1'+1)}
	[\delta_{\el_3 \el_3'} \delta_{\el_4 \el_4'} +
	 (-1)^{\Sigma_L} \delta_{\el_3 \el_4'} \delta_{\el_4 \el_3'}]
	C_{\el_1} C_{\el_1'} C_{\el_3} C_{\el_4}\,,
\label{eqn:paircov}
\end{equation}
other pairings can be found through the permutation properties
of $Q$ (or $T$).  No fundamentally new terms are introduced if three or four
$\el$'s are equal but each set of possible internal pairings in the primed
and unprimed sets must be separately accounted for.	


The total signal-to-noise for each $L$ in the four-point spectrum is 
\begin{eqnarray}
\left({S \over N }\right)^2 &\equiv&
\sum_{\el_1 \el_2 \el_3 \el_4 L}
\sum_{\el_1' \el_2' \el_3' \el_4' L'}
\langle \hat T^{\el_1 \el_2*}_{\el_3 \el_4}(L) \rangle \,
[{\rm Cov}^{-1}]
\langle T^{\el_1' \el_2'}_{\el_3' \el_4'}(L')\rangle  
\nonumber\\ &\approx& \quad  \sum_L \sum_{\el_1>\el_2>\el_3>\el_4}
{1 \over 2L +1} {| \hat T^{\el_1\el_2}_{\el_3 \el_4}(L)|^2 \over 
C_{\el_1}^\tot C_{\el_2}^\tot C_{\el_3}^\tot C_{\el_4}^\tot}\,,
\label{eqn:sn}
\end{eqnarray}
where ``Cov$^{-1}$'' indicates the matrix inverse, with elements
labelled by their configuration $(\el_1, \el_2, \el_3, \el_4, L)$,
of the covariance in 
Eqn.~(\ref{eqn:covariance}).
In the second line, the restricted sum eliminates the $4!$=24 redundant permutations
above and neglects the signal-to-noise contributed when the
the $\el$'s are equal.
In the high signal-to-noise regime, one must also include the sample 
variance of the signal.  On a cut-sky, the considerations of Appendix
\ref{sec:flatsky} imply that the overall $(S/N)^2$ is reduced by
a factor of $f_{\rm sky}$.

Note that if the sum in Eqn.~(\ref{eqn:sn}) is not restricted 
the covariance supplied by the alternate pair orderings in 
Eqn.~(\ref{eqn:pairsymmetry}) necessarily contains off diagonal terms 
that mix $L$ and $L'$.  The covariance is distributed
across many $L$'s and can lead to overestimates of the signal-to-noise
in 4-point related statistics by a factor of  $\sqrt{3}$.

\subsection{Power Spectrum Covariance}
\label{sec:pscovariance}

The trispectrum can affect two-point or power spectrum statistics by
introducing a covariance between the estimators.
The covariance of power spectrum estimators averaged over $m$ is given by
\begin{eqnarray}
    [{\rm Cov}]_{\el_{1} \el_{2}} &=&
    {1 \over 2\el_{1}+1} {1 \over 2\el_{2}+1}
    \sum_{m_{1} m_{2}} 
    \langle \Theta_{\el_{1} m_{1}} \Theta_{\el_{1} m_{1}}^{*}
    \Theta_{\el_{2} m_{2}} \Theta_{\el_{2} m_{2}}^{*}\rangle 
    - \langle \hat C_{\el_{1}} \rangle \langle \hat C_{\el_{2}}\rangle 
	\nonumber\\
   &=&{1 \over \sqrt{2\el_{1}+1}} {1 \over \sqrt{2\el_{2}+1}}
    (-1)^{\el_1+\el_2} Q^{\el_1 \el_1}_{\el_2 \el_2}(0) - 
	\langle \hat C_{\el_1} \rangle
	\langle \hat C_{\el_2} \rangle \,.
\label{eqn:pcov}
\end{eqnarray}
The expression for the covariance can be further broken into its
Gaussian and non-Gaussian pieces
\begin{eqnarray}
    [{\rm Cov}]_{\el_{1} \el_{2}}
   &=& {2 \over 2\el_{1}+1} C_{\el_{1}}^{2} \delta_{\el_{1}\el_{2}}
\nonumber\\
&& \quad	+ { (-1)^{\el_1+\el_2} 
	\over \sqrt{(2\el_1+1)({2 \el_2+1})}}
\left[
	P^{\el_1\el_1}_{\el_2\el_2}(0) + 
	{2 \over \sqrt{(2\el_1+1)(2 \el_2+1)}}
	\sum_{L=|\el_1-\el_2|}^{\el_1+\el_2}
	(-1)^L P^{\el_1 \el_2}_{\el_1 \el_2}(L) \right]\,.
\label{eqn:powercov}
\end{eqnarray}
The effect of covariance for the signal-to-noise for the estimation of
a set of underlying cosmological parameters $p_i$ 
can be calculated through the Fisher matrix
\begin{eqnarray}
F_{ij} = \sum_{\el_1 \el_2} 
{\partial C_{\el_1} \over \partial p_i} 
[{\rm Cov}^{-1}]
{\partial C_{\el_2} \over \partial p_j} \,,
\label{eqn:fishercov}
\end{eqnarray}
where ``Cov$^{-1}$'' indicates the matrix inverse of the covariance in
Eqn.~(\ref{eqn:powercov}).
In particular, if the only parameter of interest is the overall amplitude
$A$
of a known template shape, then $F_{AA} = (S/N)^2$ (see \cite{CooHu00a}).

\section{Power Spectra of Quadratic Statistics}
\label{sec:collapsedtri}

Measuring all of the configurations of the trispectrum or four-point
function is a daunting challenge.  In this section we consider 
statistics based on the identification of points in pairs
in the four-point function.  These quadratic statistics may be optimized in 
signal-to-noise for their power spectra 
by filtering the original temperature field.
We begin with general definitions for the quadratic fields in harmonic space
(\S \ref{sec:general}) and continue through a discussion of filters
(\S \ref{sec:filters}) to a consideration of specific quadratic statistics
(\S \ref{sec:temptemp}-\ref{sec:div}) and related cubic
statistics (\S \ref{sec:cubic}).  The specific statistic
and filter set that optimizes the signal-to-noise will depend on the
configuration dependence of the trispectrum signal that is to be extracted.

\subsection{General Definitions}
\label{sec:collapsed}

To probe various aspects of the trispectrum, we can form the
two-point or power spectrum statistics of a quadratic combination of
the underlying field.  
To enhance the signal-to-noise,
we begin by filtering the fields before collapsing the configuration,
\begin{equation}
\Theta^a (\bn) = \sum_{\el m} \Theta_{\el m} f^{a}_\el \, Y_\el^m(\bn)\,,
\label{eqn:filter}
\end{equation}
where the index $a=1,4$ to allow for 4 independent filters on the
fields.  
In general, the identification of points in pairs implies that 
each pair ($ab$) involves
a quadratic combination of the filtered field which in turn involves
a mode coupling sum of the harmonic coefficients
\begin{eqnarray}
x_{L M}^{ab} &=& (-1)^{M}\sum_{\el_1 m_1} \sum_{\el_2 m_2} 
		x_{\el_{1} \el_{2}}^{ab}(L)
		\Theta_{\el_1 m_1}
		\Theta_{\el_2 m_2}\sqrt{2 L+1 \over 4\pi}
		\wj{\el_1}{\el_2}{L}{m_1}{m_2}{-M} \,,
\label{eqn:generalstat}
\end{eqnarray}
where 
\begin{eqnarray}
		x_{\el_{1} \el_{2}}^{ab}(L) = f^a_{\el_1} f^b_{\el_2} 
	\sqrt{(2\el_1+1)(2\el_2+1)} x_{\el_{1}\el_{2}}(L)\,,
\end{eqnarray}
and $x_{\el_1 \el_2}(L)$ represents different weights 
for different statistics $x$ as specified below.
The power spectra statistics relating two general quadratic statistics
$x$ and $\tilde x$ may be separated into the non-Gaussian signal and
Gaussian noise as
\begin{eqnarray}
\langle x_{LM}^{12*} \tilde x_{L'M'}^{34} \rangle &=& \delta_{L L'} \delta_{M M'}
(C_L^{x\tilde x} +N_L^{x\tilde x})\,,
\end{eqnarray}
where
\begin{eqnarray}
C_L^{x\tilde x} &=& {1 \over 4\pi}{1\over 2 L+1}\sum_{\el_1\el_2\el_3\el_4}
              x_{\el_{1}\el_{2}}^{12*}(L) \tilde x_{\el_{3}\el_{4}}^{34}(L)
	(-1)^{\el_1 + \el_2 + L}
		T^{\el_1 \el_2}_{\el_3 \el_4}(L)\,,
\label{eqn:ngpower}
\end{eqnarray}
and the Gaussian noise is
\begin{eqnarray}
    N_{L}^{x\tilde x} &=&     {1 \over 4\pi} \sum_{\el_{1}\el_{2}} 
    x_{\el_{1}\el_{2}}^{12*}(L) 
    \left[ \tilde x_{\el_{1}\el_{2}}^{34}(L) + 
	(-1)^{\el_1 + \el_2 + L}
    \tilde x_{\el_{2}\el_{1}}^{34}(L) \right]
     C^\tot_{\el_1} C^\tot_{\el_2}\,.
\label{eqn:gnoise}
\end{eqnarray}
It will be useful in the following discussion of noise variance to
also define the following two auxiliary power spectra,
\begin{eqnarray}
    V_{L}^{x\tilde x (12)} &=&     {1 \over 4\pi} \sum_{\el_{1}\el_{2}} 
    x_{\el_{1}\el_{2}}^{12*}(L) 
    \left[ \tilde x_{\el_{1}\el_{2}}^{12}(L) + 
	(-1)^{\el_1 + \el_2 + L}
    \tilde x_{\el_{2}\el_{1}}^{12}(L) \right]
     C^\tot_{\el_1} C^\tot_{\el_2}\,, \nonumber\\
    V_{L}^{x\tilde x (34)} &=&     {1 \over 4\pi} \sum_{\el_{1}\el_{2}} 
    x_{\el_{1}\el_{2}}^{34*}(L) 
    \left[ \tilde x_{\el_{1}\el_{2}}^{34}(L) + 
	(-1)^{\el_1 + \el_2 + L}
    \tilde x_{\el_{2}\el_{1}}^{34}(L) \right]
     C^\tot_{\el_1} C^\tot_{\el_2}\,.
\label{eqn:vnoise}
\end{eqnarray}

The signal-to-noise ratio in this power spectrum statistic can be calculated
from equation (\ref{eqn:covariance})
for the covariance of the trispectrum,
\begin{eqnarray}
\langle \hat C_L^{x\tilde x} \hat C_L^{x'\tilde x'} \rangle 
&\approx  & {1 \over {2 L+1}}
	\left[ 
\langle V_L^{x x' (12)} \rangle
\langle V_L^{\tilde x \tilde x' (34)} \rangle 
+ 
\langle N_L^{x \tilde x'} \rangle
\langle N_L^{\tilde x x'} \rangle \right]
\,, 
\label{eqn:covtwopoint}
\end{eqnarray}
such that
\begin{eqnarray}
\left( {S \over N} \right)^2 \approx \sum_{L x\tilde x x' \tilde x'} 
{
\langle C^{x\tilde x}_L  \rangle
\langle C^{x' \tilde x'}_L \rangle
	\over \langle \hat C_L^{x\tilde x} \hat C_L^{x' \tilde x'}
	 \rangle }
\,.
\label{eqn:sntwopoint}
\end{eqnarray}
Strictly speaking, this is an inequality since we have neglected
the covariance between $L$'s dictated by the trispectrum covariance
Eqn.~(\ref{eqn:covariance}).  
Since the trispectrum covariance is distributed broadly in the allowed
$L$'s, this signal-to-noise estimate is reasonable if we restrict
the range of interest in $L$ to a small fraction of the allowed range.
The covariance can at most reduce the total signal-to-noise by
a factor of $\sqrt{3}$ for the 3 unique pairings in the trispectrum.

\subsection{Filters}
\label{sec:filters}

The filters and specific form of the statistic $x$ can be chosen
to eliminate Gaussian noise bias and/or maximize the signal-to-noise.
If ($f^1_\el$, $f^2_\el$) and ($f^{3}_\el$, $f^{4}_\el$) 
do not overlap in $\el$, then the Gaussian noise bias of Eqn.~(\ref{eqn:gnoise}) vanishes.
Furthermore, trispectrum covariance between differing $L$'s
is identically zero and Eqn.~(\ref{eqn:sntwopoint}) becomes a
strict equality.

For example the two filters may be band limited in mutually 
exclusive bands or parity limited
\begin{eqnarray}
\Theta_e(\bn) &\equiv& {1 \over 2}\left[ \Theta(\bn) + \Theta(-\bn)\right]\,,
\nonumber\\
f^{1}_\el = f^{2}_\el &\equiv& \cases { 
		1,             &\quad $\el = $even\,,  \cr
		0,            &\quad $\el = $odd\,,  }
\\
\Theta_o(\bn) &\equiv& {1 \over 2}\left[ \Theta(\bn) - \Theta(-\bn)\right]\,,
\nonumber\\
f^{3}_\el = f^{4}_\el &\equiv& \cases { 
		0,             &\quad $\el = $even\,,  \cr
		1,            &\quad $\el = $odd\,.  }
\end{eqnarray}
This choice does not eliminate the auxiliary variance power spectra in
Eqn.~(\ref{eqn:vnoise}) and more generally does not maximize the total
signal-to-noise.
Only if the filters are equal in pairs 
$f^1_\el=f^3_\el$ and $f^2_\el=f^4_\el$ are the noise and auxiliary 
variance power spectra equal such that
\begin{eqnarray}
\langle \hat C_L^{x\tilde x} \hat C_L^{x'\tilde x'} \rangle 
\approx {1 \over 2L+1} \left( 
\langle N_L^{x x'} \rangle
\langle N_L^{\tilde x \tilde x'} \rangle +
\langle N_L^{x \tilde x'} \rangle
\langle N_L^{\tilde x x'} \rangle \right)\,,
\end{eqnarray}
becomes the familiar form for the variance of the power spectra of
a set of Gaussian random fields $x$.
 
A comparison of the signal-to-noise in the full trispectrum
Eqn.~(\ref{eqn:sn}) and in a particular quadratic statistic $x$
Eqn.~(\ref{eqn:sntwopoint})
shows that the latter approaches the former if
\begin{eqnarray}
x_{\el_1\el_2}^{12}(L) x_{\el_3\el_4}^{34}(L) \rightarrow
	w(L)
	{T^{\el_1 \el_2}_{\el_3 \el_4}(L) \over 
	C^\tot_{\el_1}
	C^\tot_{\el_2}
	C^\tot_{\el_3}
	C^\tot_{\el_4}}\,,
\label{eqn:optimalweights}
\end{eqnarray}
where $w(L)$ is an arbitrary function of $L$.  To the extent that
the right hand side is factorable in $\el_{a}$, $a=1,4$ the
filter functions $f^a_{\el}$ can by chosen construct this optimal
statistic.  Since the trispectrum is in general not factorable,
we will next consider a wide range of choices for the quadratic
$x$-statistic which
can be used to construct optimal statistics for various types of
trispectrum signals.

\subsection{Temperature-Temperature}
\label{sec:temptemp}

The simplest quadratic statistic that we can form is the
product of the filtered temperature field itself,
\begin{eqnarray}
	\Theta^a(\bn) \Theta^b(\bn) 
&\equiv& 
	s^{ab}(\bn) =
\sum_{LM} s_{LM}^{ab} Y_L^M(\bn)\,,
\end{eqnarray} 
where $s_{LM}^{ab}$ is given by the general prescription of
Eqn.~(\ref{eqn:generalstat}) with $x=s$ and the weighting
\begin{equation}
	s_{\el_1\el_2}(L) = \wj{\el_{1}}{\el_{2}}{L}{0}{0}{0}\,,
\quad \even\,.
\end{equation}
``Even'' denotes the fact that $s$ selects out $\el_1+\el_2+L =$even 
by virtue of the Wigner-$3j$ symbol.
The non-Gaussian power spectrum $C_L^{ss}$
is then given by Eqn.~(\ref{eqn:ngpower}) in terms of the trispectrum.
The total signal-to-noise of this statistic can be estimated by
retaining just the $x=x'=\tilde x=\tilde x'=s$ terms in 
Eqn.~(\ref{eqn:sntwopoint}).

\subsection{Temperature-Gradient}
\label{sec:tempgrad}

The product of the filtered temperature field and the gradient of
the filtered temperature field probes another aspect the 
trispectrum.  This product is a vector field on the sky and
may be broken up into components as
\begin{equation}
\Theta^a (\bn) \nabla_i \Theta^b (\bn)
\equiv \sum_{\pm}{1 \over \sqrt{2}} [\alpha_1 \pm i \alpha_2]^{ab}({\bn})
	{1 \over \sqrt{2}}(\hat{\bf e}_\phi \mp \hat{\bf e}_\theta)_i\,.
\end{equation}
The components $\alpha_1\pm i \alpha_2$ are spin-1 objects that can be decomposed
in the spin-1 spherical harmonics \cite{spinylm},
\begin{eqnarray}
 \left[\alpha_1 \pm i \alpha_2 \right]^{ab}(\bn) &=&  \sum_{L M} 
(c \pm ig)_{L M}^{ab}\, {}_{\pm 1} Y_L^M (\bn)\,,
\end{eqnarray}
where $c$ and $g$ are the multipole analogues of the curl and gradient pieces.
These quadratic statistics again follow the general form of
Eqn.~(\ref{eqn:generalstat}) with $x = c$, $g$ and weightings
\begin{eqnarray}
    c_{\el_{1}\el_{2}}(L) &\equiv& -\sqrt{\el_{2}(\el_{2}+1)}
    \wj{\el_{1}}{\el_{2}}{L}{0}{-1}{1}\,, \quad \odd
\,,\\
    g_{\el_{1}\el_{2}}(L) &\equiv& i\sqrt{\el_{2}(\el_{2}+1)}
    \wj{\el_{1}}{\el_{2}}{L}{0}{-1}{1}\,, \quad \even\,.
\end{eqnarray}
Here and below ``even'' (``odd'') denotes the fact that the 
expression holds for $\el_1+\el_2+L=$even (odd) and vanishes otherwise.
If the trispectrum is parity invariant 
(zero if $\el_1+\el_2+\el_3+\el_4$ =odd),
the cross power spectra $C_L^{gc} = 0 = C_L^{sc}$ vanish. 
The remaining power spectra and their covariance
are described by the general forms of Eqn.~(\ref{eqn:ngpower}),
(\ref{eqn:gnoise}), and (\ref{eqn:covtwopoint}).

\subsection{Gradient-Gradient}
\label{sec:gradgrad}

The product of temperature gradients can in general be decomposed
into three quadratic statistics 
\begin{eqnarray}
[\nabla_i \Theta_{a} (\bn)]\, [\nabla_j \Theta_{b} (\bn)]
       &\equiv& t^{ab} (\bn) g_{ij}(\bn) + \sum_{\pm}
	[q \pm i u]^{ab}(\bn) {\bf \sigma}_{ij}^{\pm}(\bn)
	+ v(\bn) \epsilon_{ij}(\bn)\,,
\end{eqnarray}
where $g_{ij}$ is the metric on the 2-sphere,
\begin{equation}
\sigma^{\pm}_{ij}(\bn) = {1 \over 2} 
(\hat {\bf e}_\theta \mp \hat {\bf e}_\phi)_i
(\hat {\bf e}_\theta \mp \hat {\bf e}_\phi)_j\,,
\end{equation}
gives the basis for a trace-free
symmetric tensor field on the sky, and
\begin{equation}
\epsilon_{ij}(\bn) = 
	({\bf e}_\theta)_j
	({\bf e}_\phi)_i
-
	({\bf e}_\theta)_i
	({\bf e}_\phi)_j 
\end{equation}
gives the basis for a trace-free antisymmetric tensor field on the sky.
The flat-sky versions of
these statistics were first employed by \cite{ZalSel99} for CMB lensing
and note that $q$, $u$, $v$ are analogous to the similarly named Stokes
parameters for polarization.

As is the case for the CMB polarization, these three fields 
may be decomposed into multipole moments of the
spherical harmonics and spin-2 spherical harmonics \cite{spinylm},
\begin{eqnarray}
 t^{ab}(\bn) &=& \sum_{LM}  t_{LM}^{ab} Y_L^M (\bn)\,, \nonumber\\
 v^{ab}(\bn) &=& \sum_{LM}  v_{LM}^{ab} Y_L^M (\bn)\,, \nonumber\\
 \left[q\pm i u \right]^{ab}(\bn) &=&  \sum_{L M} (e \pm ib)_{L M}^{ab} \,{}_{\pm 2} Y_L^M (\bn)\,,
\end{eqnarray}
where the moments follow the general prescription of Eqn.~(\ref{eqn:generalstat}) 
with $x=t$, $e$, $b$, $v$ and weights
\begin{eqnarray}
    t_{\el_{1}\el_{2}}(L) &\equiv& - {1\over 2}\sqrt{\el_{1}(\el_{1}+1)}
    \sqrt{\el_{2}(\el_{2}+1)}
    \wj{\el_{1}}{\el_{2}}{L}{-1}{1}{0}\,, \quad \even\,,
    \nonumber\\
    v_{\el_{1}\el_{2}}(L) &\equiv&  {i\over 2}\sqrt{\el_{1}(\el_{1}+1)}
    \sqrt{\el_{2}(\el_{2}+1)} 
    \wj{\el_{1}}{\el_{2}}{L}{-1}{1}{0}\,, \quad \odd\,,
    \nonumber\\
    e_{\el_{1}\el_{2}}(L) &\equiv& {1 \over 2}\sqrt{\el_{1}(\el_{1}+1)}
    \sqrt{\el_{2}(\el_{2}+1)}
    \wj{\el_{1}}{\el_{2}}{L}{-1}{-1}{2}\,, \quad \even\,,
    \nonumber\\
    b_{\el_{1}\el_{2}}(L) &\equiv& -{i \over 2}\sqrt{\el_{1}(\el_{1}+1)
}
    \sqrt{\el_{2}(\el_{2}+1)}
    \wj{\el_{1}}{\el_{2}}{L}{-1}{-1}{2}\,, \quad\odd\,.
\end{eqnarray}
If the trispectrum is parity invariant, cross power spectra
exist only among ($t$, $e$, $g$, $s$) and ($\beta$,$v$,$c$).
These power spectra and their covariance
again are described by the general forms of Eqn.~(\ref{eqn:ngpower}),
(\ref{eqn:gnoise}), and (\ref{eqn:covtwopoint}).

\subsection{Temperature-Hessian}
\label{sec:temphess}

Similarly to the gradient-gradient case, the product of the temperature
and the second derivatives or Hessian of the temperature field can be
decomposed into three quadratic statistics
\begin{eqnarray}
\Theta^a (\bn) \nabla_i \nabla_j \Theta^b (\bn)
       &\equiv& h^{ab} (\bn) g_{ij}(\bn) + \sum_{\pm}
	[\eta_1 \pm i \eta_2]^{ab}(\bn) {\bf \sigma}_{ij}^{\pm}(\bn)\,,
\end{eqnarray}
which themselves may be decomposed into multipole moments of the
spherical harmonics and spin-2 spherical harmonics,
\begin{eqnarray}
 h^{ab}(\bn) &=& \sum_{LM}  h_{LM}^{ab} Y_L^M (\bn) \,,\nonumber\\
 \left[\eta_1+i \eta_2 \right]^{ab}(\bn) &=&  \sum_{L M} (\epsilon + i\beta)_{L M}^{ab} \,{}_{\pm 2} Y_L^M (\bn)\,,
\end{eqnarray}
where the moments follow the general prescription of Eqn.~(\ref{eqn:generalstat}) 
with $x=h$, $\epsilon$, $\beta$ and weights
\begin{eqnarray}
    h_{\el_{1}\el_{2}}(L) &\equiv& -{1 \over 2}\el_2(\el_2+1)
    \wj{\el_{1}}{\el_{2}}{L}{0}{0}{0}\,, \quad \even\,,
    \nonumber\\
&=& -{1 \over 2} \el_2(\el_2+1) s_{\el_1\el_2} \,,\nonumber\\
    \epsilon_{\el_{1}\el_{2}}(L) &\equiv& {1 \over 2} 
    \sqrt{(\el_2+2)! \over (\el_2-2)!}
    \wj{\el_{1}}{\el_{2}}{L}{0}{-2}{2} \,, \quad \even\,,
    \nonumber\\
    \beta_{\el_{1}\el_{2}}(L) &\equiv& -{i\over 2} 
    \sqrt{(\el_2+2)! \over (\el_2-2)!}
    \wj{\el_{1}}{\el_{2}}{L}{0}{-2}{2} \,, \quad \odd\,,
\end{eqnarray}
Again parity invariance requires that power spectra exist only between
($h$, $\epsilon$, $t$, $e$, $s$) and ($\beta$,$b$,$v$,$c$).
Likewise the general formula for power spectra and their covariance
again apply.

\subsection{Temperature-Temperature Hessian}
\label{sec:temptemphess}

Auxiliary two-point statistics can be formed from the fundamental ones
above. For example
\begin{eqnarray}
\nabla_i \nabla_j [\Theta^a (\bn) \Theta^b (\bn)]
       &\equiv& \tilde t^{ab}(\bn)
	g_{ij}(\bn) + 
	\sum_{\pm}
	[\tilde q \pm i \tilde u]^{ab} (\bn) \sigma_{ij}^\pm (\bn)\,,
\end{eqnarray}
and
\begin{eqnarray}
 \tilde t^{ab}(\bn) &=& \sum_{LM}  \tilde t_{LM}^{ab} Y_L^M (\bn)\,, \nonumber\\
 \left[\tilde q+i \tilde u \right]^{ab}(\bn) &=&  \sum_{L M} (\tilde e + i\tilde b)_{L M}^{ab} \,{}_{\pm 2} Y_L^M (\bn)\,,
\end{eqnarray}
imply that
\begin{eqnarray}
\tilde t_{\el_1\el_2} 
		      &=& t_{\el_1\el_2}+t_{\el_2\el_1} + h_{\el_1 \el_2}
			+ h_{\el_2 \el_1} 
			= -{L(L+1)\over 2} 
			\wj{\el_1}{\el_{2}}{L}{0}{0}{0}\,,
\nonumber\\
\tilde e_{\el_1\el_2} &=& e_{\el_1\el_2}+e_{\el_2\el_1} 
		+ \epsilon_{\el_1\el_2} + \epsilon_{\el_2\el_1} 
			={1 \over 2}\sqrt{(L+2)! \over (L-2)!}
			\wj{\el_1}{\el_{2}}{L}{0}{0}{0}\,,
\nonumber\\
\tilde b_{\el_1\el_2} &=& b_{\el_1\el_2}+b_{\el_2\el_1} 
		+ \beta_{\el_1\el_2} + \beta_{\el_2\el_1} 
			=0\,.\nonumber
\end{eqnarray}			
Again power spectra follow from the general relations. 

\subsection{Temperature-Gradient Divergence}
\label{sec:div}

The divergence of the temperature-gradient field of
\S \ref{sec:tempgrad} is also an auxiliary statistic
\begin{equation}
\nabla^i [
\Theta^a (\bn) \nabla_i \Theta^b (\bn)]
\equiv d^{ab} ({\bn}) Y_L^M (\bn) \nonumber\,,\\
\end{equation}
with
\begin{equation}
d^{ab}(\bn) = \sum_{LM}  d_{LM}^{ab} Y_L^M (\bn)\,. \nonumber\\
\end{equation}
The weights are related to the others as
\begin{eqnarray}
d_{\el_1 \el_2} &=& 
	\sqrt{L(L+1) \el_2(\el_2+1)} 
    \wj{\el_{1}}{\el_{2}}{L}{0}{-1}{1} \,, \quad \even\,, \nonumber\\	
	&=& -i \sqrt{L(L+1)}  g_{\el_1 \el_2}
	= 2t_{\el_1 \el_2} +2 h_{\el_1 \el_2}\,.
\end{eqnarray}
Again power spectra follow from the general relations. 

\subsection{Cubic Statistics}
\label{sec:cubic}
Finally the cross correlation of cubic statistics with linear statistics
are also related to the quadratic statistics introduced above.
For example 
\begin{equation}
\langle  \Theta^1(\bn_1) \Theta^2(\bn_1) 
\Theta^3(\bn_1) \,\Theta^4(\bn_2)  \rangle =
\sum_{\el m} (-1)^m [C_\el^{ss(3)}+N_\el^{ss(3)}] Y_\el^{-m}(\bn_1) Y_\el^m(\bn_2)
\,.\nonumber\\
\end{equation}
More generally, the cubic power spectra corresponding to the various
$x$-statistics are given by
\begin{eqnarray}
C_\el^{x\tilde x(3)} &=& 
		\sum_{\el_1\el_2\el_3 L}
		{1 \over 4\pi}{1\over 2 L+1}
              x_{\el_{1}\el_{2}}^{12*}(L) \tilde x_{\el_{3}\el}^{34}(L)
		(-1)^{\el_1+\el_2+L}
		T^{\el_1 \el_2}_{\el_3 \el}(L)\,,
\label{eqn:ng3power}
\end{eqnarray}
with Gaussian noise bias
\begin{eqnarray}
    N_{\el}^{x\tilde x(3)} &=&     {1 \over 4\pi} \sum_{\el_{1}L} 
    x_{\el_{1}\el}^{12*}(L) 
    \left[ \tilde x_{\el_{1}\el}^{34}(L) + 
		(-1)^{\el_1+\el_2+L}
    \tilde x_{\el\el_{1}}^{34}(L) \right]
     C^\tot_{\el_1} C^\tot_{\el}\,.
\label{eqn:g3noise}
\end{eqnarray}
and auxiliary noise variance
\begin{eqnarray}
    V_{\el}^{x\tilde x (3,12)} &=&     {1 \over 4\pi} \sum_{\el_{1}L} 
    x_{\el_{1}\el}^{12*}(L) 
    \left[ \tilde x_{\el_{1}\el}^{12}(L) + 
		(-1)^{\el_1+\el_2+L}
    \tilde x_{\el \el_{1}}^{12}(L) \right]
     C^\tot_{\el_1} C^\tot_{\el}\,, \nonumber\\
    V_{\el}^{x\tilde x (3,34)} &=&     {1 \over 4\pi} \sum_{\el_{1}L} 
    x_{\el_{1}\el}^{34*}(L) 
    \left[ \tilde x_{\el_{1}\el}^{34}(L) + 
		(-1)^{\el_1+\el_2+L}
    \tilde x_{\el\el_{1}}^{34}(L) \right]
     C^\tot_{\el_1} C^\tot_{\el}\,,
\label{eqn:v3noise}
\end{eqnarray}
i.e. the multipole index of the power spectra is no longer the
diagonal of the trispectrum configuration but rather one of its sides.
These cubic statistics thus probe a different projection of the
trispectrum information but are based on the same set of filters and employ
the same statistical formalism.

\section{CMB Lensing}
\label{sec:lensing}

In this section, we consider the trispectrum signal generated
by the weak gravitational lensing of the CMB temperature anisotropies by the
large-scale structure in the Universe.  In \S \ref{sec:lensingtri}
we derive the full trispectrum for lensing and relate it to
the underlying deflection (or convergence) power spectrum.
Zaldarriaga \cite{Zal00} previously considered the lensing 
trispectrum in the small-scale or flat-sky approximation.  
The gravitational lensing effect is known to be dominated by potential
fluctuations on the largest scales where an all-sky treatment of the 
trispectrum is desirable \cite{Hu00b}.  
In \S \ref{sec:lensingsn}, we show that the signal-to-noise 
in the trispectrum is both large and highly configuration dependent
for experiments that can resolve multipole moments $\el \agt 1000$.
The divergence statistic introduced in \S \ref{sec:div} is shown in
\S \ref{sec:lensingdiv} to be optimal for measuring 
the underlying deflection power
spectrum at its peak at low multipoles.  It benefits from substantially
higher signal-to-noise as compared with the gradient-gradient 
quadratic statistics introduced by Zaldarriaga \& Seljak \cite{ZalSel99}.
Finally we consider tests for the robustness of the divergence statistic
with differing filter sets in \S \ref{sec:robustness} and the
degradation in power spectrum estimation from lensing covariance in
\S \ref{sec:lensingcov}.

\begin{figure*}[tb]
\centerline{\epsfxsize=4truein\epsffile{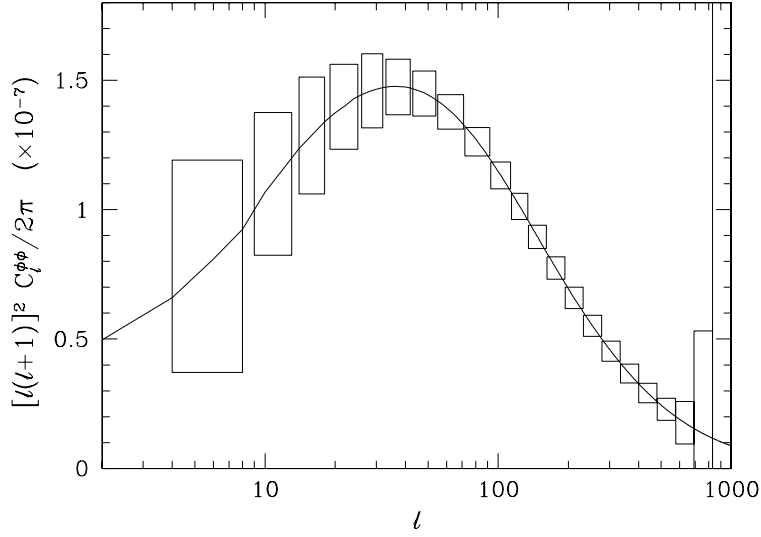}}
\caption{The power spectrum of the deflection angle in the fiducial
$\Lambda$CDM model.  Errors boxes represent the $1 \sigma$ 
errors from Gaussian noise on the divergence statistic binned 
in the bands shown.  
The divergence estimator of Eqn.~(\protect\ref{eqn:divweight})-(\protect\ref{eqn:divfilter})
 is optimal for the low multipoles and
reduces the variance in the power spectrum estimation by more
than an order of magnitude as compared with the
gradient-gradient statistics of \protect\cite{ZalSel99}.}
\label{fig:div}
\end{figure*}

\subsection{General Trispectrum}
\label{sec:lensingtri}

We begin by briefly reviewing the effect of gravitational lensing on
the harmonics of the CMB temperature field and refer the reader to
\cite{Hu00b} for details of its calculation in a given cosmology.
For reference, we employ the same $\Lambda$CDM cosmology used
there, with parameters $\Omega_m=0.35$, $\Omega_\Lambda=0.65$, $h=0.65$,
$n=1$ and $\delta_H=4.2 \times 10^{-5}$.

Weak lensing of the CMB remaps the primary anisotropy according
to the deflection angle $\nabla\phi(\bn)$
\begin{eqnarray}
\Theta(\bn) & = &  \tilde \Theta(\bn + \nabla\phi) \nonumber\\
        & = &
\tilde \Theta(\bn) + \nabla_i \phi(\bn) \nabla^i \tilde \Theta(\bn) 
+\ldots\,,
\end{eqnarray}
where the tilde represents the unlensed field and $\ldots$ represent higher
order terms in the Taylor expansion. The lensing potential field
$\phi(\bn)$ is a lensing-probability weighted projection of the
Newtonian potential along the line of sight [see \cite{Hu00b}, Eqn.~(21)].

\begin{figure*}[tb]
\centerline{\epsfxsize=4truein\epsffile{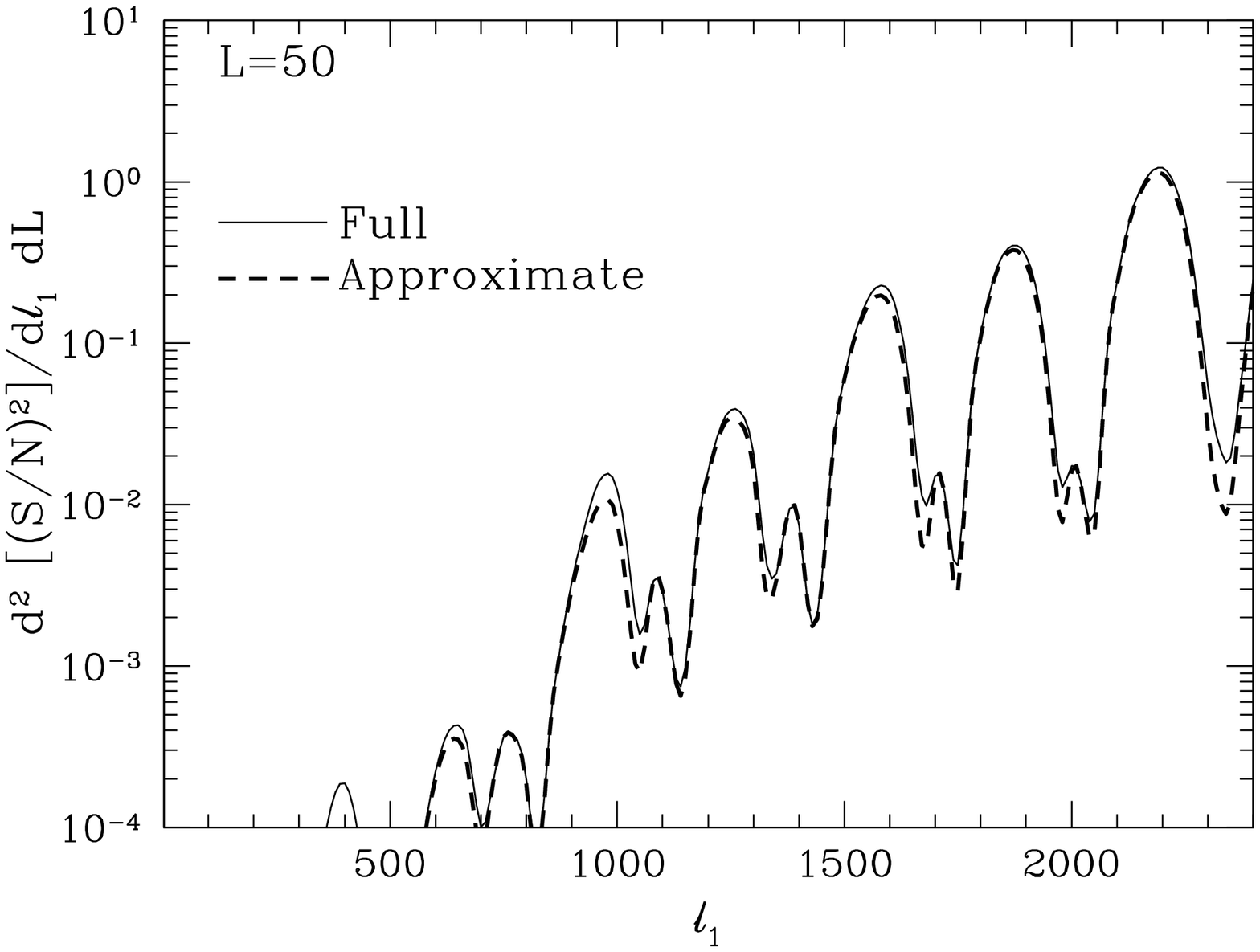}}
\caption{Contributions to the $(S/N)^2$ from trispectra configurations
with a fixed diagonal $L$ and maximum side length $\el_1$, summed over
the remaining three sides.  Solid lines represent the full calculation
of the trispectrum terms; dashed lines represent the pairwise approximation
of Eqn.~(\protect\ref{eqn:trispectapprox}).  The signal-to-noise in the low $L$ 
trispectrum is highly dependent on the configuration.}
\label{fig:snlmax}
\end{figure*}

The spherical harmonic coefficients of the lensed CMB temperature field
become
\begin{eqnarray}
\Theta_{\el m} &\approx& \tilde \Theta_{\el m} + \int d\bn Y_\el^{m*}(\bn)
        \nabla_i \phi(\bn) \nabla^i \tilde \Theta(\bn) + \ldots \nonumber\\
&=& \tilde
 \Theta_{\el m} + \sum_{L M} \sum_{\el' m'}
                        \phi_{L M}  \tilde \Theta_{\el' m'} 
(-1)^m \wj{\el}{\el'}{L}{m}{-m'}{-M} F_{\el L \el'} + \ldots\,,
\label{eqn:thetalm}
\end{eqnarray}
where
\begin{eqnarray}
\sqrt{ 4\pi \over (2\el+1)(2\el'+1)(2L+1)}
F_{\el L \el'} 
&=& 
{1 \over 2}
[L(L+1)+\el'(\el'+1)-\el(\el+1)]     
\wj{\el}{\el'}{L}{0}{0}{0}\,, \nonumber\\
 &=& -\sqrt{L(L+1) \el'(\el'+1)} 
\wj{\el}{\el'}{L}{0}{-1}{1}, \quad \even\,,
\end{eqnarray}
where recall that  ``even'' denotes the fact that only
$\el+\el'+L$=even is non-vanishing.
Gravitational lensing generates a change in the power spectrum
that has been well studied \cite{Sel96,ZalSel98,MetSil97,Hu00b}.  It produces two changes to the 4 point
function.  The first is that the unlensed $\tilde C_{\el}$ in the Gaussian
4-point contribution must be replaced with the lensed $C_{\el}$. 
The second is that it generates a trispectrum an underlying reduced
form [see Eqn.~(\ref{eqn:reducedtrispect})] 
of 
\begin{eqnarray}
\trispect^{\el_1 \el_2}_{\el_3 \el_4}(L) &=& C_{L}^{\phi\phi} 
\tilde C_{\el_{2}} \tilde C_{\el_{4}} F_{\el_{1}L\el_{2}}
F_{\el_{3}L\el_{4}} \,.
\label{eqn:lenstrispect}
\end{eqnarray}
Note the geometric interpretation: the lensing generates a 
trispectrum or 
quadrilateral configuration of $\el_{1} \ldots \el_{4}$ where one of
the diagonals is supported by the lensing potential power spectrum
$C_{L}^{\phi\phi}$.  Note that the power spectrum of
the deflection field is given by $L(L+1) C_L^{\phi\phi}$ and is
the fundamental quantity of interest.  It is plotted in Fig.~\ref{fig:div}
for the fiducial $\Lambda$CDM cosmology.  It is important to note
that most of the power in the deflections is coming from a rather large
scale or low multipole $L \approx 50$.  Contrast this with the more familiar
convergence power spectrum $C_L^{\kappa\kappa} = [L(L+1)/2]^2 C_L^{\phi\phi}$
which peaks at much smaller angular scales.

\begin{figure*}[tb]
\centerline{\epsfxsize=4truein\epsffile{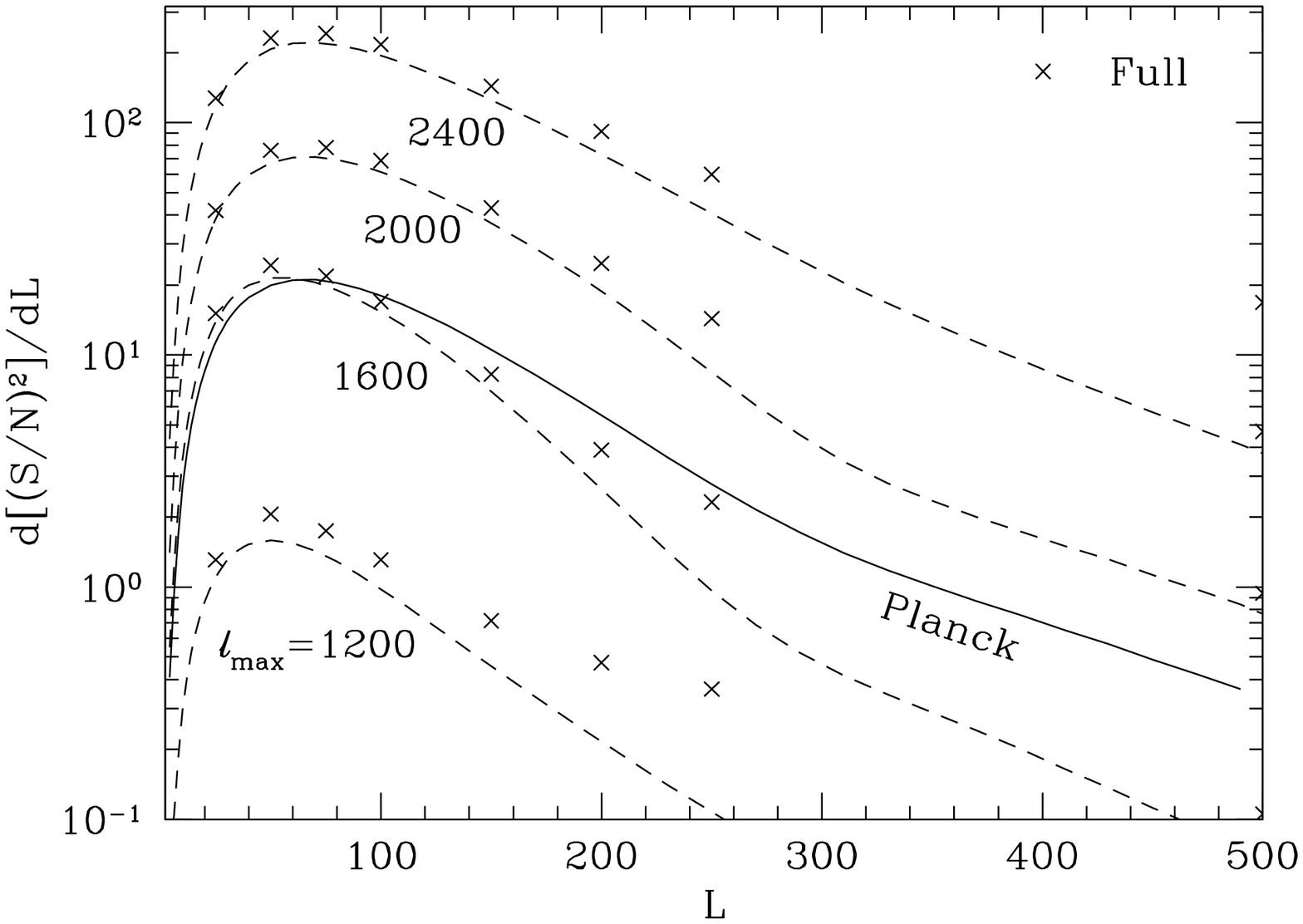}}
\caption{Cumulative signal to noise in the trispectra configurations with
the diagonal $L$ summed over $\el_1\ldots \el_4$.  Dashed lines represent
an ideal experiment where $C_\el=C_\el^\tot$ out to a
 maximum $\el=\el_{\rm max}$ 
; solid lines represent the Planck experiment.
Lines represent the 
approximation of Eqn.~(\protect\ref{eqn:trispectapprox}); 
points represent the calculation using the full trispectrum for an ideal
experiment.}
\label{fig:snl}
\end{figure*}

\subsection{Total Signal-to-Noise}
\label{sec:lensingsn}

From the considerations of \S \ref{sec:signaltonoise}, we can calculate
the total signal-to-noise in the trispectrum for lensing in the full-sky
formalism.  Flat-sky estimates of the total signal-to-noise have
been calculated in \cite{Zal00} (see also Appendix \ref{sec:flatsky}).
The all-sky expressions are cumbersome to calculate due to the presence of
the Wigner-6$j$ symbol that expresses the alternate recouplings of
the trispectrum $\el$'s.  We use the recursion technique outlined in
Appendix \ref{sec:w6j} for these calculations.

In Figure~\ref{fig:snlmax}, we show the signal-to-noise contributions in a given
mode $L=50$ of the trispectrum from
a given $\el_1$ (summed over $\el_2$, $\el_3$, $\el_4$) assuming an ideal
experiment $C_\el^\tot=C_\el$.  
The signal-to-noise is quite high and approaches unity per $\el_1$ mode and
$L$ mode at $\el_1 \approx 2000$.  Moreover, the contributions
as a function of $\el_1$ show striking features.  These features can be
understood by approximating the trispectrum by its fundamental pairing
($\el_1$, $\el_2$), ($\el_3$,$\el_4$) in Eqn.~(\ref{eqn:reducedtrispect})
\begin{equation}
T^{\el_1 \el_2}_{\el_3 \el_4}(L) \approx P^{\el_1 \el_2}_{\el_3 \el_4} =
	C_L^{\phi\phi}
	\left( \tilde C_{\el_2} F_{\el_1 L \el_2} + \tilde C_{\el_1} F_{\el_2 L \el_1}\right)
	\left( \tilde C_{\el_4} F_{\el_3 L \el_4} + \tilde C_{\el_3} F_{\el_4 L \el_3}\right)\,.
\label{eqn:trispectapprox}
\end{equation}
Figure~\ref{fig:snlmax} (dashed lines) verifies that this is a very good
approximation for the range of interest
$\el_1 \gg L$.  The reason is that these configurations represent
flattenned quadrilaterals where one diagonal is much greater than the other.
Since lensing effects peak at low $L$, the other pairings are correspondingly
suppressed.  These properties are hidden in the real-space 4-point function
and highlights the benefit of considering harmonic-space statistics.

To the extent that $\tilde C_{\el}$ is constant, the two terms within each set of
parenthesis in Eqn.~(\ref{eqn:trispectapprox}) cancel.  Thus, the trispectrum
picks out features in the underlying unlensed power spectrum, specifically
those due to the acoustic peaks in the power spectrum.  Note that
the effects on the power spectrum itself exhibits the same effect: lensing
acts to smooth the acoustic features in the spectrum.  This structure implies
that optimizing the $\el$-range of filters in the quadratic statistics is
important for maximizing the signal-to-noise.  

The cumulative signal-to-noise integrating out to $\el_1=\el_{\rm max}$ as
a function of $L$ in an ideal experiment is shown in Fig.~\ref{fig:snl}.
The approximation of 
Eqn.~(\ref{eqn:trispectapprox}) begins to break down as $L$ approaches
$\el_{\rm max}$ but always in the sense that it {\it underestimates} the
total signal-to-noise (dashed lines vs. points).  This breakdown occurs since the two
diagonals of a trispectrum quadrilateral become comparable and either
can be supported by the lensing power in Eqn.~(\ref{eqn:lenstrispect}).
We also show in Fig.~\ref{fig:snl} (solid lines) the cumulative signal-to-noise
for the Planck satellite \cite{Planck} with $C_\el^\tot$ taken from \cite{CooHu00a}.  
Planck approximates an ideal experiment with $\el_{\rm max} \approx 1600$.

\begin{figure*}[tb]
\centerline{\epsfxsize=4truein\epsffile{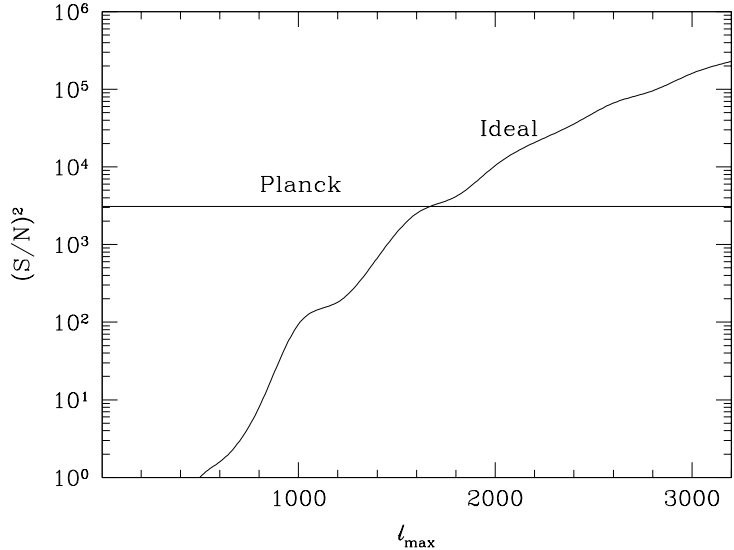}}
\caption{Approximate total $(S/N)^2$ in the trispectrum for 
an ideal experiment out to
$\el=\el_{\rm max}$ and the Planck experiment.  The Planck experiment 
approximates an ideal experiment of $\el \approx 1600$ with a $(S/N)^2
\sim 3100$.}
\label{fig:sntot}
\end{figure*}

Finally, under the approximation of Eqn.~(\ref{eqn:trispectapprox}) which
slightly underestimates the total signal-to-noise, we can plot the
cumulative signal-to-noise summed over all $L$ 
as a function of $\el_{\rm max}$
(see Fig.~\ref{fig:sntot}).  Again an $\el_{\rm max}\approx 1600$
approximates the Planck experiment whose total $(S/N)^2 \approx 3100$.

\subsection{Divergence Statistic}
\label{sec:lensingdiv}

The structure in the signal-to-noise curves imply that it is important
to select a quadratic statistic that captures this structure.  
Following the considerations of \S \ref{sec:collapsed},  we can search for
the quadratic statistic that optimizes the signal-to-noise
ratio in the power spectrum of the deflection.  
Recall that in general the multipole moments of the quadratic
statistic with filters $f_\el^a$ and $f_\el^b$
\begin{eqnarray}
x_{L M}^{ab} &=& (-1)^{M}\sum_{\el_1 m_1} \sum_{\el_2 m_2} 
		x_{\el_{1} \el_{2}}^{ab}(L)
		\Theta_{\el_1 m_1}
		\Theta_{\el_2 m_2}\sqrt{2 L+1 \over 4\pi}
		\wj{\el_1}{\el_2}{L}{m_1}{m_2}{-M} \,,
\label{eqn:generalstat2}
\end{eqnarray}
are defined in terms of the weight function $x_{\el_1 \el_2}^{ab}$.
Under the approximation for the 
trispectrum of Eqn.~(\ref{eqn:trispectapprox}),
Eqn.~(\ref{eqn:optimalweights}) gives the optimal weights as
\begin{equation}
x_{\el_1 \el_2}^{12} \propto { \tilde C_{\el_2} F_{\el_1 L \el_2}
		+ \tilde C_{\el_1} F_{\el_2 L \el_1} \over 
		C_{\el_1}^\tot
		C_{\el_2}^\tot\,}.
\end{equation}
Since the lensing trispectrum is symmetric in 
$\el_1 \leftrightarrow \el_2$, the temperature-gradient divergence
statistic whose weights are
\begin{eqnarray}
x_{\el_1 \el_2}^{12} &=& {1 \over 2}  
	(d_{\el_1 \el_2}^{12}  + d_{\el_2 \el_1}^{12})\,, \nonumber\\
d_{\el_1 \el_2}^{12} &=&
	 f_{\el_1}^1 f_{\el_2}^2 F_{\el_1 L \el_2} \sqrt{ 4\pi \over 2L+1}\,,
\label{eqn:divweight}
\end{eqnarray}
is optimal if the underlying temperature field is first filtered with
\begin{eqnarray}
f^1_\el = f^3_\el &=& - {A \over C_\el^\tot} \nonumber\,,\\
f^2_\el = f^4_\el &=&   {\tilde C_\el \over C_\el^\tot}\,.
\label{eqn:divfilter}
\end{eqnarray}

We choose the proportionality constant to return the properly normalized
deflection power spectrum
\begin{equation}
A =  \sqrt{L(L+1)} (2L+1)\left[ \sum_{\el_1 \el_2} {
	(\tilde C_{\el_2} F_{\el_1 L \el_2} +
	\tilde C_{\el_1} F_{\el_2 L \el_1} )^2
	  \over 2 C_{\el_1}^\tot C_{\el_2}^\tot\,} \right]^{-1}\,.
\end{equation}
The signal-to-noise in this statistic is maximal
under the assumption that the trispectrum can be approximated as 
Eqn.~(\ref{eqn:trispectapprox}) as is the case for $L \alt$ several
hundred. 
The total signal-to-noise as calculated from
Eqn.~(\ref{eqn:sntwopoint}) for Planck is formally $(S/N)^2 \approx 4050$.
Compare this with the gradient-gradient statistic of \cite{ZalSel99};
for the unfiltered $e^{ab}({\bn})\equiv {\cal E}({\bn)}$ the
$(S/N)^2 \approx 135$.  

The total signal-to-noise is allowed to exceed the maximum 
estimate of the previous section
since the latter is strictly a lower limit.  However the 
underlying approximation that the estimates at all $L$'s
are independent breaks down when integrating over a wide range in $L$ causing
a small reduction in the number of independent modes
(see \S \ref{sec:collapsed}).

The divergence statistic may also be used as a direct estimator
of the deflection (or equivalently the convergence) field itself.
Generalizing the argument of \cite{ZalSel99},
one can think of the quadratic statistic $x$ in  Eqn.~(\ref{eqn:generalstat2})
as averaging over many independent (high-$\el$  or small-scale) realizations
of the unlensed CMB field with a fixed realization of the large-scale
deflection field
\begin{eqnarray}
\langle x_{L M}^{ab} \rangle_{\rm CMB}
	&=& \phi_{LM} \sqrt{1 \over 4\pi (2L+1)}
		\sum_{\el_1\el_2} x_{\el_1\el_2}^{ab}
    \left( \tilde C_{\el_1} F_{\el_2 L \el_1} + \tilde C_{\el_2} F_{\el_1 L \el_2}\right)\,,
\quad (L>0)\,.
\end{eqnarray}
For the divergence statistic filtered as in Eqn.~(\ref{eqn:divfilter})
all contributions add coherently so that,
\begin{eqnarray}
\langle d_{L M}^{ab} \rangle_{\rm CMB} &=&
	 {\phi_{LM} \over (2L+1)}
		\sum_{\el_1\el_2} {A \over 2 C_\el^\tot C_\el^\tot} 
    \left( \tilde C_{\el_1} F_{\el_2 L \el_1} + \tilde C_{\el_2} F_{\el_1 L \el_2}\right)^2\,\nonumber\\
	&=& \sqrt{L(L+1)} \phi_{LM} \,.
\label{eqn:deflectionest}
\end{eqnarray}
From the multipole moments one can reconstruct the deflection or 
convergence map \cite{Hu01b}.  Of course the fact that we average over only a 
finite number of independent modes of the primary anisotropy means that
the resulting map will be noisy, with noise properties given by
the Gaussian noise power spectra.

\subsection{Robustness Tests}
\label{sec:robustness}

The divergence statistic contains enough signal-to-noise for Planck that
the data may be further subdivided to check for robustness of the
statistic.  Especially worrying is the possibility that galactic and
extragalactic foregrounds and systematic effects might generate a false
signal.  Even if these contaminants contribute only Gaussian noise, 
one must
subtract out the noise bias from the power spectrum estimators
with the filters defined in Eqn.~(\ref{eqn:divfilter}), 

Recalling the discussion of the filters in \S \ref{sec:filters}, we
can eliminate Gaussian noise bias as well as noise-correlation between
differing $L$ by defining non-overlapping filters sets 
($f^1_\el$, $f^2_\el$) and ($f^3_\el$, $f^4_\el$).   The resulting
estimates of the deflection field $d_{LM}^{12}$ and $d_{LM}^{34}$ would
then have statistically independent Gaussian noise properties such
that the noise bias is eliminated in their cross correlation.  
Furthermore, if the signal is really due to lensing the various estimates
of the deflection power spectrum must agree within their errors. 

The price of dividing up the sample in this way is the signal-to-noise
in any given set.  For example,  by band-limiting the filters of
Eqn.~(\ref{eqn:divfilter}) to $500 < \el < 1400$ for the $(12)$ set
and $\el > 1400$ for the $(34)$ set, the total signal-to-noise is 
reduced by $\sim \sqrt{2}$ and correspondingly the errors in 
Fig.~\ref{fig:div} are increased by $\sim \sqrt{2}$.
Note that in this case, the underlying filtered temperature maps 
contain no power in the multipoles of interest $L\sim 100$.
Such a scheme would still yield a highly significant detection and
help protect against contaminants.  Filtered versions of other
quadratic statistics can also serve as consistency checks.  For
example, even the simple temperature-temperature $s$ statistic of
\S \ref{sec:temptemp}, yields a $(S/N)^2 \sim 200$ once it is 
is filtered according to the peaks in the Fig.~\ref{fig:snlmax}.

\begin{figure*}[tb]
\centerline{\epsfxsize=4truein\epsffile{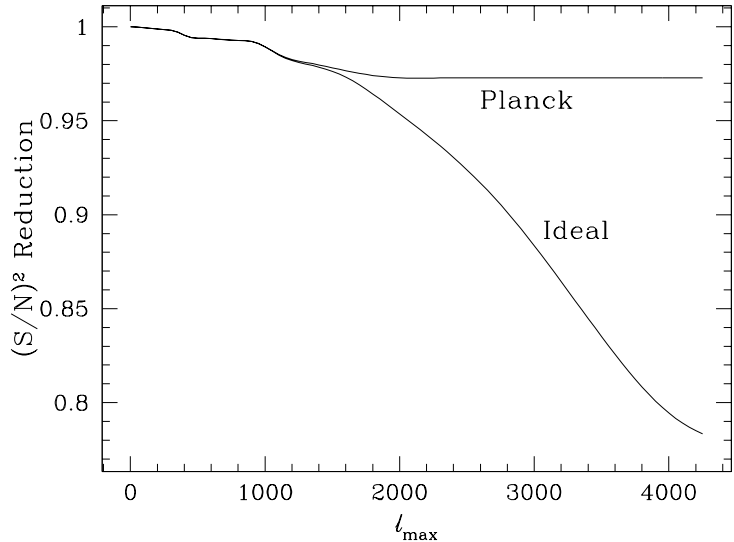}}
\caption{Degradation in the total $(S/N)^2$ in the power spectrum
due to covariance from gravitational lensing.   The degradation is
minimal for the Planck experiment or any that is cosmic variance
limited only out to $\el \sim 2000$.}
\label{fig:cov}
\end{figure*}

\subsection{Power Spectrum Covariance}
\label{sec:lensingcov}

One might worry that the high signal-to-noise in the trispectrum for lensing
comes at the expense of degraded signal to noise in the power spectrum
due to covariance between the estimators.  
Fortunately this is not the case in the $\el \alt 1000$ regime of the
acoustic peaks.  
From Eqn.~(\ref{eqn:pcov})-(\ref{eqn:fishercov}), we can calculate
the degradation in the total $(S/N)^2$.   This degradation is
shown in Fig.~\ref{fig:cov} for the Planck satellite and 
an ideal experiment out to $\el_{\rm max}$.   

\section{Discussion}
\label{sec:discussion}

We have provided a systematic study of the angular trispectrum or
four-point function of the CMB temperature field.  Symmetry
considerations dictate the fundamental form of the trispectrum
and govern the Gaussian noise properties of its estimators.  
The large number of independent configurations of the trispectrum imply that
even subtle physical effects may be detectable when all of the information 
in the trispectrum is brought to bear.

In practice, extracting all of the information in the trispectrum will
be a difficult computational task.  We have thus also 
conducted a systematic study of the power spectra of quadratic
statistics which probe different aspects of the trispectrum.  
Techniques developed
for extracting power spectrum statistics from large data sets can then
be brought to bear on the four-point function.  The drawback is
that this compression of information is in general a lossy procedure.
We have therefore examined a wide range of quadratic statistics
and prefiltering schemes.  Given a target form for the trispectrum signal,
these statistics can be optimized in their signal-to-noise for the
power spectra.

As an example, we have reconsidered the four-point correlation generated
by weak lensing of the primary anisotropies by large-scale structure in
the Universe as a means of recovering the power spectrum of the
deflection angles or convergence \cite{ZalSel99}.  
CMB weak lensing provides a unique
probe of the large-scale properties of the convergence field.
We identify a specific quadratic statistic, the divergence of the
temperature-weighted gradient field, that achieves the
maximal signal-to-noise in this limit.  For the Planck satellite,
the total $(S/N)^2 \approx 4000$ and represents a reduction in 
the noise variance
on the convergence power spectrum by over an order of magnitude as
compared with the gradient-gradient estimators of \cite{ZalSel99}.  
There is sufficient signal-to-noise to conduct filtering tests to
eliminate noise-bias and check for consistency between multipole 
subsets of the data.  We plan to explore further the properties 
of these estimators and their use in extracting cosmological 
parameters in a separate work.

The lensing example illustrates the importance of examining the 
configuration properties of the trispectrum when designing 
statistical estimators based on the four-point function.  Extracting the wealth of information potentially buried in
the trispectrum will be a rich field for future studies.

\medskip 
\noindent
{\it Acknowledgments:} I thank A.R. Cooray and M. Zaldarriaga for
many useful discussions.  This work was supported by NASA ATP
NAG5-10840 and
a Sloan Foundation Fellowship.
\appendix
\section{Wigner-6$j$ Symbol}
\label{sec:w6j}

\subsection{Useful Properties}

The Wigner-6$j$ symbol expresses the relationship between two
distinct couplings of three angular momenta
\begin{eqnarray}
{\bf J_d} &=& {\bf J_a} + {\bf J_b} + {\bf J_c} \nonumber\\
          &=& {\bf J_e} + {\bf J_c} \nonumber\\
	  &=& {\bf J_a} + {\bf J_f}\,,
\end{eqnarray}
such that the eigenstates of the ($ec$) coupling are 
related to the eigenstates of the ($af$) coupling as
\begin{eqnarray}
    |(ec)d\gamma\rangle 
    =
    |(af)d\gamma\rangle
    \sqrt{(2 e+1)(2f +1)} (-1)^{\Sigma}
    \wsixj{a}{b}{e}{c}{d}{f}\,,
\end{eqnarray}
where $\Sigma \equiv a+b+c+d$.  Geometrically, the Wigner-6$j$
represents a quadrilateral with sides ($a$,$b$,$c$,$d$) 
whose diagonals form the triangles ($a$,$d$,$f$), ($b,c,f$),
($c$,$d$,$e$), ($a,b,e$) or the three dimensional
tetrahedron composed of these four triangles.  It vanishes if the any of the
triplets fail to satisfy the triangle rule. The symmetries
are related to rotations of the tetrahedron that interchange
the vertices.  The result is that the symbol is invariant
under the interchange of any two columns and under the interchange
of the upper and lower arguments in any two columns.

The Wigner-6$j$ symbol can thus be used to permute the pairings
in a set of Wigner-$3j$ symbols
\begin{eqnarray}
    &&\sum_{f} (2 f+1)(-1)^{\Sigma+f-e-\alpha-\gamma}
    \wsixj{a}{b}{e}{c}{d}{f}
    \wj{b}{c}{f}{\beta}{\gamma}{-\phi}
    \wj{a}{f}{d}{\alpha}{\phi}{-\delta}
    =
    \wj{a}{b}{e}{\alpha}{\beta}{-\epsilon}
    \wj{e}{c}{d}{\epsilon}{\gamma}{-\delta}\,,
\end{eqnarray}
or equivalently by the orthogonality relation of the Wigner-$3j$ symbols
\begin{eqnarray}
    \wsixj{a}{b}{e}{c}{d}{f}
    &=& \sum_{\alpha\beta\gamma} \sum_{\delta \epsilon \phi}
    (-1)^{{e+f+\epsilon+\phi}}
    \wj{a}{b}{e}{\alpha}{\beta}{\epsilon}
    \wj{c}{d}{e}{\gamma}{\delta}{-\epsilon}
    \wj{a}{d}{f}{\alpha}{\delta}{-\phi}	
    \wj{c}{b}{f}{\gamma}{\beta}{\phi}\,.
\label{eqn:w3jw6j}
\end{eqnarray}
Finally, the Wigner-$6$j symbol obeys 
\begin{equation}
\sum_e (2e+1) 
\wsixj{a}{b}{e}{c}{d}{f} 
\wsixj{a}{b}{e}{c}{d}{g} = {\delta_{f g} \over 2 f+1}\,,
\label{eqn:normalization}
\end{equation}
and
\begin{equation}
\sum_e (-1)^{e + f + g} (2 e+1)
\wsixj{a}{b}{e}{c}{d}{f} 
\wsixj{a}{b}{e}{d}{c}{g} = 
\wsixj{a}{c}{g}{h}{d}{f} \,.
\label{eqn:recoupling}
\end{equation}

\subsection{Evaluation}

Closed form expressions exist for special cases of the arguments.  
For example,
\begin{eqnarray}
    \wsixj{a}{b}{e}{c}{d}{0} = {(-1)^{{a+b+e}} \over
    \sqrt{(2 a+1)(2 b+1)}} {\delta_{a,d}\delta_{b,c}}\,.
\end{eqnarray}
More generally, they may be computed 
efficiently by a recursive algorithm introduced by \cite{SchGor75}.  
Let us define
\begin{eqnarray}
h(j_1) = \wsixj{j_1}{j_2}{j_3}{\el_1}{\el_2}{\el_3}\,.
\end{eqnarray}
The $h(j_1)$ satisfy the recursion
\begin{equation}
j_1 E(j_1+1) h(j_1+1) + F(j_1) h(j_1) + (j_1+1) E(j_1) h(j_1-1) = 0 \,,
\end{equation}
where
\begin{eqnarray}
E(j_1)&=&\Big\{ \left[
			j_1^2 - (j_2-j_3)^2 
 	       \right]
               \left[
		       (j_2 + j_3 + 1)^2  - j_1^2
	       \right]
	       \left[  
		       j_1^2 - (\el_2 - \el_3)^2 
	       \right] 
		 \left[ (\el_2 + \el_3 + 1)^2 - j_1^2 \right]\Big\}^{1/2}
	\nonumber\\
F(j_1)&=&(2 j_1+1) \Big\{
			    j_1(j_1+1)[-j_1(j_1+1) + j_2(j_2+1) + j_3(j_3+1)]
		\nonumber\\ && \quad
			   +\el_2(\el_2+1) 
			    [j_1(j_1+1) + j_2(j_2+1) - j_3(j_3+1)]
		\nonumber\\ && \quad
			   + \el_3(\el_3+1)
			    [j_1(j_1+1) - j_2(j_2+1) + j_3(j_3+1)]
		\nonumber\\ && \quad
			  -2 j_1(j_1+1) \el_1 (\el_1+1) \Big\}
\end{eqnarray}
For a stable recursion, one begins at both of the two ends
$j_{1 \rm min}= {\rm max}(|j_2 - j_3|,|\el_2 -\el_3|)$
$j_{1 \rm max}= {\rm min}(j_2 + j_3,\el_2 +\el_3)$
with the boundary conditions $E(j_{\rm 1 min})=0$ and $E(j_{\rm 1 max}+1)=0$
and matches the two in the middle.  

The normalization is fixed by 
\begin{eqnarray}
\sum_{j_1} (2 j_1+1)(2 \el_1+1) [h(j_1)]^2 &=& 1\,,\nonumber\\
{\rm sgn} [h(j_{1 \rm max})] &=& (-1)^{j_2 + j_3 + \el_2 + \el_3}\,,
\end{eqnarray}
which follow from Eqn.~(\ref{eqn:normalization}).

\section{Flat Sky Approximation}
\label{sec:flatsky}

In the flat sky approximation, one decomposes the temperature field
into Fourier harmonics
\begin{equation}
\langle \Theta(\bn_1) \ldots \Theta(\bn_n) \rangle =
\int {d^2 l_1 \over (2\pi)^2}  \ldots
\int {d^2 l_n \over (2\pi)^2}  
\langle \Theta(\bl_1)\ldots \Theta(\bl_n) \rangle
e^{i \bl \cdot \bn_1} \ldots e^{i \bl \cdot \bn_2}\,.
\end{equation}
Statistical isotropy is enforced by demanding that the correlation function
be invariant under an arbitrary translation and rotation in the plane.  
Parity invariance is enforced by demanding symmetry under inversion of the
coordinates or reflection across one of the coordinate axes.

As usual translational symmetry $\bn_i \rightarrow \bn_i + {\bf C}$, 
where ${\bf C}$ is a constant vector, is enforced by the closure condition that
the $n$-point function is proportional to 
\begin{equation}
(2\pi)^2 \delta(\bl_1 + \ldots + \bl_n)\,.
\end{equation}
The wavevectors $\bl_i$ thus form a geometric figure of $n$, possibly intersecting
sides.
Rotational invariance for the two-point function 
and rotational and parity invariance for the three-point function
imply that the corresponding harmonic spectra are functions only of
the lengths of the sides:
\begin{eqnarray}
\langle \Theta(\bl_1) \Theta(\bl_2) \rangle &=& 
(2\pi)^2 \delta(\bl_{12}) C_{(l_1)}\,, \nonumber\\
\langle \Theta(\bl_1) \Theta(\bl_2) \Theta(\bl_3) \rangle &=& 
(2\pi)^2 \delta(\bl_{123}) B_{(l_1,l_2,l_3)} \,,
\end{eqnarray}
where ${\bl}_{i\ldots j} \equiv \bl_i +\ldots + \bl_j$ and $B$ should be symmetric
against permutations of its arguments.  
For the four-point function, rotational and parity invariance implies that
the quadrilateral formed by the four wavevectors is a function of the lengths
of the sides and the lengths of the two diagonals,
\begin{eqnarray}
\langle \Theta(\bl_1) \ldots  \Theta(\bl_4) \rangle &=& 
(2\pi)^4 \left[
\delta(\bl_{12})\delta(\bl_{34}) C_{\el_1} C_{\el_2} 
+
\delta(\bl_{13})\delta(\bl_{24}) C_{\el_1} C_{\el_3}
+
\delta(\bl_{14})\delta(\bl_{23}) C_{\el_1} C_{\el_4}
\right] 
\nonumber\\
&& \quad +
(2\pi)^2 \delta(\bl_{1234}) T^{(l_1,l_2)}_{(l_3,l_4)}(l_{12},l_{13}) 
\,.
\label{eqn:flatfourpoint}
\end{eqnarray}
To parallel our treatment of the all-sky four-point function let us
break $Q$ into its three distinct pairings and demand symmetry with respect
to permutation of the arguments
\begin{eqnarray}
T^{(l_1,l_2)}_{(l_3,l_4)}
 =
  P^{(l_1,l_2)}_{(l_3,l_4)}(l_{12}) +
  P^{(l_1,l_3)}_{(l_2,l_4)}(l_{13}) +
  P^{(l_1,l_4)}_{(l_3,l_2)}(l_{14})  \,.
\label{eqn:flatqp}
\end{eqnarray}
Note that $l_{14}$ is a function of the
other two diagonals.
$P$ is symmetric under interchange of its upper and lower arguments
as well as ordering within them.
%

The relationship between the all sky and flat sky spectra can be obtained
by noting that \cite{Hu00b}
\begin{eqnarray}
\Theta_{l m} &=& 
	i^m \sqrt{ 2 l+1 \over 4\pi} \int {d \phi_\bl \over 2\pi}
	e^{i m\phi_\bl} \Theta(\bl)\,, \nonumber\\
\delta(\bl_{i \ldots j}) & = & 
\int {d \bn \over (2\pi)^2} e^{i \bl_{i \ldots j} \cdot \bn} \,,\nonumber\\
e^{i \bl \cdot \bn} & =& \sqrt{2 \pi \over l} \sum_m i^m Y_l^m (\bn) e^{i m \phi_{\bl}}\,,
\end{eqnarray}
where $\phi_\bl$ is the polar angle of $\bl$.  It is a straightforward
exercise to show that
\begin{eqnarray}
C_l &=& C_{(l)}\,, \nonumber\\
B_{l_1 l_2 l_3} &=& \wj{l_1}{l_2}{l_3}{0}{0}{0}
		  \sqrt{(2l_1+1)(2l_2+1)(2l_3+1) \over 4\pi}
		   B_{(l_1,l_2,l_3)}\,.
\end{eqnarray}
For the trispectrum, we begin with the general correspondence
\begin{eqnarray}
\langle \Theta_{l_1 m_1} \ldots \Theta_{l_4 m_4} \rangle_c
 &=& \left( \prod_{i=1,4} \sqrt{l_1 \over 2\pi} \int{d\phi_{\bf l_i} \over 2\pi}
	     e^{-i m_i \phi_{\bl_i}}\right) 
	(2\pi)^2 \delta(\bl_{1234})  T^{(l_1,l_2)}_{(l_3,l_4)}(l_{12})\,,
\end{eqnarray}
where ``$c$'' denotes the subtraction of the Gaussian piece in
Eqn.~(\ref{eqn:flatfourpoint}).
We then exploit the pair symmetry of the trispectrum exhibited in
Eqn.~(\ref{eqn:flatqp})
by breaking the delta function into
the corresponding pairs.  For the $(12)$, $(34)$ pairing,
\begin{eqnarray}
\delta(\bl_{1234}) &=& \int {d^2 L}\,
			\delta(\bl_1+\bl_2+{\bf L})
			\delta(\bl_3+\bl_4-{\bf L})\,, 
\end{eqnarray}
such that $L = l_{12}$.  Expanding the delta functions in spherical harmonics,
\begin{eqnarray}
\delta(\bl_{1234}) &=& 
			{1  \over (2\pi)^2}
			  \sum_{m_1\ldots  m_4} \sum_{L M} 
			\left( \prod_{i = 1,4} \sqrt{2 \pi \over l_i}
				e^{i m_i \phi_{\bl_i}} \right) 
			{2 L+1 \over 4\pi}
			\nonumber\\ && \times 
				\sqrt{(2l_1+1)\ldots (2l_4+1)} 
				\wj{l_1}{l_2}{L}{0}{0}{0}
				\wj{l_3}{l_4}{L}{0}{0}{0}
			  \nonumber\\
			 &&\times 
				(-1)^M 
				\wj{l_1}{l_2}{L}{m_1}{m_2}{M}
				\wj{l_3}{l_4}{L}{m_3}{m_4}{-M}\,.
\end{eqnarray}
Substituting back in and integrating over the polar angles, we obtain
the general correspondence
\begin{eqnarray}
P^{l_1 l_2}_{l_3 l_4}(L) &= & 
				{2L+1 \over 4\pi}
				\sqrt{(2l_1+1)\ldots(2l_4+1)} 
				\wj{l_1}{l_2}{L}{0}{0}{0}
				\wj{l_3}{l_4}{L}{0}{0}{0} 
				P^{(l_1,l_2)}_{(l_3,l_4)}(L)\,,
\end{eqnarray} 
from which we can construct the relation for $T^{l_1 l_2}_{l_3 l_4}$.

We can now also make the correspondence between the signal-to-noise
in the all-sky and flat-sky formalisms.  The weighting of four-point
terms that maximizes the signal to noise is \cite{Zal00}
\begin{equation}
\left( {S \over N} \right)_{\rm tot}^2 = 
		{f_{\rm sky} \over \pi} {1 \over 24}
		{1 \over (2\pi)^4} \int d^2 \bl_1 \int d^2 \bl_2 \int d^2 \bl_3
		\int d^2 \bl_4 \delta({\bl_{1234}})
		{ | T^{(l_1,l_2)}_{(l_3,l_4)}|^2  \over 
			C_{l_1}^\tot
			C_{l_2}^\tot
			C_{l_3}^\tot
			C_{l_4}^\tot}\,,
\end{equation}
where $f_{\rm sky}$ is the fraction of sky covered by the sample.

The square of $T$ in this expression contains cross term in the $P$ pairings.
\begin{eqnarray}
\left( {S \over N} \right)_{\rm tot}^2& = &
\left( {S \over N} \right)_{\rm (12,12)}^2 
+ \left( {S \over N} \right)_{\rm (13,13)}^2 
+ \left( {S \over N} \right)_{\rm (14,14)}^2 
+ 2\left( {S \over N} \right)_{\rm (12,13)}^2 
+ 2\left( {S \over N} \right)_{\rm (12,14)}^2 
+ 2\left( {S \over N} \right)_{\rm (13,14)}^2 \,.
\end{eqnarray}
The correspondence with the all-sky expression Eqn.~(\ref{eqn:sn}) 
can be established
by considering the terms pair-by-pair.
For example for the (12,13)
 term, one expands the delta function
in the $(12)$, $(34)$ pairing as above and inserts an additional delta 
function 
\begin{equation}
\pi \delta (\bl_{1234}) = \pi \int d^2 L_{13} 
		\delta(\bl_1+\bl_3-{\bf L}_{13})
		\delta(\bl_2+\bl_4+{\bf L}_{13})\,,
\end{equation}
with the understanding that $\delta({\bf 0}) = V/(2\pi)^2 = 1/\pi$.
Expanding the delta functions in spherical harmonics we can integrate
over azimuthal angles to obtain
\begin{eqnarray}
\left( {S \over N} \right)^2_{(12,13)} 
&=& {f_{\rm sky} \over 24} \sum_{\el_1 m_1 \ldots \el_4 m_4}
\sum_{L_{12} M_{12}}
\sum_{L_{13} M_{13}}
(-1)^{M_{12}+M_{13}} 
\wj{\el_1}{\el_2}{L_{12}}{m_1}{m_2}{M_{12}}
\wj{\el_3}{\el_4}{L_{12}}{m_3}{m_4}{-M_{12}}
\nonumber\\&& \quad \times
\wj{\el_1}{\el_3}{L_{13}}{m_1}{m_3}{M_{13}}
\wj{\el_2}{\el_4}{L_{13}}{m_2}{m_4}{-M_{13}}
{P^{\el_1\el_2*}_{\el_3\el_4}(L_{12}) 
P^{\el_1\el_3}_{\el_2\el_4}(L_{13})  \over
			C_{l_1}^\tot
			C_{l_2}^\tot
			C_{l_3}^\tot
			C_{l_4}^\tot} \,,
\nonumber\\
&=& {f_{\rm sky} \over 24} \sum_{\el_1 \ldots \el_4} \sum_{L_{12} L_{13}}
        (-1)^{\el_2+\el_3}\wsixj{\el_1}{\el_2}{L_{12}}{\el_4}{\el_3}{L_{13}}
	{P^{\el_1\el_2*}_{\el_3\el_4} (L_{12})
	P^{\el_1\el_3}_{\el_2\el_4} (L_{13}) \over
			C_{l_1}^\tot
			C_{l_2}^\tot
			C_{l_3}^\tot
			C_{l_4}^\tot} \,,
\end{eqnarray}
where we have used Eqn.~(\ref{eqn:w3jw6j}) to rewrite the sum over the Wigner-$3j$ symbols
in terms of the $6j$-symbol.  
Proceeding similarly for all terms in the
signal-to-noise expression, 
we obtain
\begin{equation}
\left({S \over N }\right)^2_{\rm tot} = {f_{\rm sky} \over 24}
\sum_{L} \sum_{\el_1 \el_2 \el_3 \el_4}
{1 \over 2L +1} {| T^{\el_1\el_2}_{\el_3 \el_4}(L)|^2 \over 
C_{\el_1}^\tot C_{\el_2}^\tot C_{\el_3}^\tot C_{\el_4}^\tot}\,,
\end{equation}
where we have employed Eqn.~(\ref{eqn:recoupling}) to reexpress the 
$(13,14)$ term.
The factor of 24 comes from the $4!$ permutations of each quadruplet in
the all-sky expression.  The $f_{\rm sky}$ term is the reduction in signal-to-noise 
due to incomplete sky coverage.


\begin{thebibliography}{99}

\mybib{Goldberg \& Spergel}{GolSpe99}
\aut{Goldberg}{D.M} \laut{Spergel}{D.N} 
\refs{The Microwave Background Bispectrum, Paper II: A Probe of the Low Redshift Universe}
{\PRD}{59}{103002}{1999}{astro-ph/9811251}

\mybib{Seljak & Zaldarriaga}{SelZal98i}
\aut{Seljak}{U} \laut{Zaldarriaga}{M} 
\refs{Direct Signature of Evolving Gravitational Potential from 
Cosmic Microwave Background}
{\PRD}{60}{043504}{1999}{astro-ph/9811123}

\mybib{Cooray \& Hu}{CooHu00a}
\aut{Cooray}{A.R}  \laut{Hu}{W}
\refs{Imprint of Reionization on the Cosmic Microwave Background 
Bispectrum}{\ApJ}{534}{533}{2000}{astro-ph/9910397}

\bibitem{early} T. Falk, R. Rangarajan, M. Srednicki, \ApJ 403, L1 (1993);
	        X. Luo, D.N. Schramm, \PRL 71, 1124 (1993);
	        A. Gangui, F. Lucchin, S. Matarrese,  S. Mollerach, \ApJ
		430, 447 (1994)

\mybib{Luo}{Luo94}
\aut{Luo}{X} 
\refs{The angular bispectrum of the cosmic microwave background}
{\ApJ}{427}{71}{1994}{astro-ph/9312004}

\mybib{Zaldarriaga}{Zal00}
\aut{Zaldarriaga}{M} 
\refs{Lensing of the CMB: Non-Gaussian Aspects}
{\PRD}{62}{063510}{2000}{astro-ph/9910498}

\mybib{Hu}{Hu00b}
\aut{Hu}{W}
\refs{Weak Lensing of the CMB: A Harmonic Approach}{\PRD}{62}{043007}{2000}{astro-ph/0001303}; the relation between $\phi$ and $\Phi$ or $\kappa$ 
has the wrong sign.


\mybib{Spergel \& Goldberg}{SpeGol99}
\aut{Spergel}{D.N}  \laut{Goldberg}{D.M}
\refs{The Microwave Background Bispectrum, Paper I: Basic Formalism}
       {\PRD}{59}{103001}{1999}{astro-ph/9811252}

\bibitem{spinylm} 
E. Newman, R. Penrose, J. Math Phys. 7, 863 (1966); 
J.N. Goldberg et al., ibid, 8, 2155 (1967).

\mybib{Zaldarriaga \& Seljak}{ZalSel99}
\aut{Zaldarriaga}{M}  \laut{Seljak}{U}
\refs{Reconstructing Projected Matter Density Power Spectrum from
       Cosmic Microwave Background}{\PRD}{59}{123507}{1999}{astro-ph/9810257}

\mybib{Seljak}{Sel96}
\aut{Seljak}{U}
\refs{Gravitational Lensing Effect on Cosmic Microwave Background Anisotropies:	A Power Spectrum Approach}{\ApJ}{463}{1}{1996}{astro-ph/9505109}

\mybib{Zaldarriaga \& Seljak}{ZalSel98}
\aut{Zaldarriaga}{M} \laut{Seljak}{U}
\refs{Gravitational Lensing Effect on Cosmic Microwave Background
Polarization}{\PRD}{58}{023003}{1998}{astro-ph/9803150} 

\mybib{Metcalf \& Silk}{MetSil97}
\aut{Metcalf}{R.B}  \laut{Silk}{J}
\refs{Gravitational Magnification of the Cosmic Microwave 
Background}{\ApJ}{489}{1}{1997}{astro-ph/9708059}

\bibitem{Planck}
http://astro.estec.esa.nl/SA-general/Projects/Planck; we take the inverse-variance weighted noise as in \cite{CooHu00a}

\mybib{Hu}{Hu01b}
\aut{Hu}{W} 
\refs{Mapping the Dark Matter with the CMB Damping Tail}
{\ApJL}{submitted}{}{2001}{astro-ph/0105424}

\bibitem{SchGor75}
K. Schulten, R. Gordon, J. Math Phys. 16, 1971 (1975)

     

\end{thebibliography}
\end{document}